\documentclass[11pt]{article}
\pdfoutput=1

\usepackage{jheppub}
\bibstyle{JHEP}

\usepackage[english]{babel} 
\usepackage[utf8]{inputenc}  
\usepackage[T1]{fontenc}
\usepackage{lmodern}
\usepackage{amsmath}         
\usepackage{amsfonts}          
\usepackage{amssymb}
\usepackage{graphicx} 
\usepackage{bbold}
\usepackage{bbm}
\usepackage[colorlinks=true,urlcolor=blue,linkcolor=blue,citecolor=blue,linktocpage=true]{hyperref}
\usepackage{color}
\usepackage{comment}
\usepackage{tikz}
\usetikzlibrary{decorations.markings}

\newcommand{\tr}{\mathrm{tr}}
\def\({\left(}
\def\){\right)}
\newcommand{\ket}[1]{| #1 \rangle }
\newcommand{\bra}[1]{\langle #1 |}

\newcommand{\figref}[1]	{{fig.~\ref{#1}}}
\newcommand{\secref}[1]	{{sec.~\ref{#1}}}

\newcommand{\Cal}[1]{{\cal #1}}

\renewcommand{\d}[2][]{\text{d}^{#1}#2}

\begin{document} 
	
	\title{Berry phases, wormholes and factorization in AdS/CFT}
	
	\author{Souvik Banerjee,}
	\author{Moritz Dorband,}
	\author{Johanna Erdmenger,}
	\author{Ren\'e Meyer and\\ {} \hspace{-3mm} }
	\author{Anna-Lena Weigel}
	
	\affiliation{Institute for Theoretical Physics and Astrophysics and W\"urzburg-Dresden Cluster of Excellence ct.qmat, Julius-Maximilians-Universit\"at W\"urzburg, Am Hubland, 97074 W\"{u}rzburg, Germany}
	
\emailAdd{\{souvik.banerjee, moritz.dorband, erdmenger, rene.meyer, anna-lena.weigel\}@physik.uni-wuerzburg.de}
	
\abstract{For two-dimensional holographic CFTs, we demonstrate the role of Berry phases for relating the non-factorization of the Hilbert space to the presence of wormholes. The wormholes are characterized by a non-exact symplectic form that gives rise to the Berry phase. For wormholes connecting two spacelike regions in gravitational spacetimes, we find that the non-exactness is linked to a variable appearing in the phase space of the boundary CFT. This variable corresponds to a loop integral in the bulk. Through this loop integral, non-factorization becomes apparent in the dual entangled CFTs. 
Furthermore, we classify Berry phases in holographic CFTs based on the type of dual bulk diffeomorphism involved. We distinguish between Virasoro, gauge and modular Berry phases, each corresponding to a spacetime wormhole geometry in the bulk. Using kinematic space, we extend a relation between the modular Hamiltonian and the Berry curvature to the finite temperature case. We find that the Berry curvature, given by the Crofton form, characterizes the topological transition of the entanglement entropy in  presence of a black hole.
}


\maketitle

\section{Introduction}
\label{sec:intro}
Within the AdS/CFT correspondence \cite{Maldacena:1997re}, there is a class of conformal field theories, namely holographic CFTs, which possess semiclassical gravity duals. Holographic CFTs can be characterized by a large central charge and a sparse spectrum of operators with low conformal dimensions whose correlators factorize \cite{El-Showk:2011yvt}. One particularly interesting example of a state in a holographic CFT is the thermofield double (TFD) state which is the maximally entangled pure state comprising energy eigenstates of two identical holographic CFTs, such that each of the entangling systems are thermal with identical temperature. While each of the thermal CFTs are holographically identified with a black hole in AdS spacetime, the entangled state is dual to an eternal black hole with a wormhole connecting the two boundaries across its interior. This idea was first advocated in \cite{VanRaamsdonk:2010pw} and later entered the ER=EPR proposal \cite{Maldacena:2013xja}. This proposal has interesting consequences - while the gravitational wormhole is a topological object that connects two spacelike separated regions of spacetime across an event horizon, its CFT dual is nothing but a surprising dynamics of entanglement between degrees of freedom of different parts of the CFT. The most surprising aspect of this connection is that while in gravity, wormholes lead to a semiclassical geometry connecting subsystems of the CFT, in the dual theory this is a very complicated dynamics of quantum information processing with apparently no classical interaction between the subsystems. These two completely different descriptions of information processing, in the bulk and at the boundary, lead to several interesting puzzles, the most recent being the factorization puzzle which makes precise the aforementioned statement in terms of the factorization of the respective Hilbert spaces. While at the boundary, due to the lack of an explicit classical interaction, the effective Hilbert space has a factorized structure, the Hilbert space in the bulk cannot be factorized in presence of the semiclassical wormhole. Such a difference in the structures of the bulk and the boundary Hilbert spaces is in apparent conflict with the AdS/CFT correspondence.

To understand the origin of and eventually solve such puzzling differences between gravitational systems and their holographic counterpart, it will be instrumental to relate the dynamics of entanglement to a topological feature in a generic quantum mechanical system. A first step toward this goal was taken in \cite{Verlinde:2021kgt}, where topological wormholes were defined in terms of a non-exact symplectic form $\Omega$ on the phase space of a given quantum mechanical system. This symplectic form appears in the thermal partition function as \cite{Verlinde:2021kgt}
\begin{align}
    Z(\beta)=\int[dX]\,\exp\left(\int_D\Omega-\oint_{\partial D}H dt\right).
    \label{eq:partition_fucntion}
\end{align}
A non-exact $\Omega$ results in non-integrable phase factors which assume a natural interpretation as Berry phases, originally introduced in \cite{Berry:1984jv}. In addition, for non-exact $\Omega$ the path integral \eqref{eq:partition_fucntion} does not factorize and thus depends on the topology of the surface $D$, leading to a topological wormhole. This relation between non-exactness of the symplectic form and topological wormholes has been studied in \cite{Nogueira:2021ngh} for a quantum mechanical two-spin model. The authors showed that even for quantum mechanical systems as simple as two coupled spins in presence of a magnetic field, a topological wormhole can be realized and characterized by the Berry phase. Also, possible experimental realizations of this two-spin model and its entanglement and topological features were proposed. Moreover, through a careful comparison of the Hilbert space of gravity to that of the quantum mechanical system, it was further established in \cite{Nogueira:2021ngh} that the Berry phase probes the entanglement structure in an analogous way both in quantum mechanics and in gravity: In the quantum mechanics model, entangled states with different Berry phase with respect to the magnetic field may nevertheless have the same entanglement entropy. A similar structure arises for wormholes connecting two boundary CFTs as in \cite{VanRaamsdonk:2010pw,Maldacena:2013xja}, due to the presence of a non-exact symplectic form originating from the singularity of the timelike Killing vector at the black hole horizon. Building up on this, it is an interesting and ongoing question how holographic setups can be brought to the laboratory; for a recent review one may consider \cite{Bhattacharyya:2021ypq}.

Motivated by these promising results in quantum mechanical systems, here we study Berry phases in two-dimensional holographic CFTs. Since these CFTs have gravitational duals, they allow us to study how the factorization properties of spacetime wormholes in the dual bulk geometry are encoded within the CFT. Similar to topological wormholes, spacetime wormholes are defined by a non-exact symplectic form leading to a non-factorized partition function \eqref{eq:partition_fucntion}, but additionally carry an interpretation as a geometric construction connecting two spacelike-separated regions in spacetime. 

For spacetime wormholes, we find that the non-exactness of the symplectic form, and thus the Berry curvature, is linked to a variable in the boundary phase space. For purely topological wormholes, no such variable is present. From the bulk perspective, this variable is either the holonomy associated to the wormhole or a time shift in one of the boundary time coordinates. The time shift is induced by the wormhole due to the absence of a global timelike Killing vector in the eternal AdS black hole geometry. We show that these variables also appear as phase space coordinates in the dual CFT. Since the CFT Berry curvature is given in terms of these variables, the non-factorization of the bulk geometry then becomes manifest also in the dual CFT. We argue that the variable may be interpreted as a coupling parameter between the dual CFTs. When this variable is set to zero, the Berry phase associated to the spacetime wormhole vanishes, indicating a decoupling of both CFTs. In some cases, we may still calculate the Berry phase in a single copy of the CFT. The Berry phase is then associated to a topological wormhole. 

Furthermore, we classify three types of Berry phases appearing in CFTs and the roles thereof in probing the entanglement structure of the Hilbert space. The first type of Berry phase is the {\bf Virasoro Berry phase}. For a single CFT,  these Berry phases originate from conformal transformations in the boundary CFT and may be defined in terms of coadjoint orbits. In particular, the Berry connection is the symplectic potential on the coadjoint orbit. These Berry phases are extensively studied from the CFT side in \cite{Oblak:2017ect}. In \cite{Compere:2015knw,Sheikh-Jabbari:2016unm}, the geometries dual to Virasoro coadjoint orbits are presented. {\it We present the holographic interpretation of Virasoro Berry phase in terms of the classification of bulk geometries via Virasoro coadjoint orbits. We show that this classification is instrumental in categorizing different kinds of wormholes arising from entangled CFTs.}
According to this classification, the Berry phases discussed in  \cite{Oblak:2017ect} arise from non-trivial conformal transformations corresponding to {\it improper diffeomorphisms} in the bulk. These diffeomorphims change the bulk geometry. Each of these bulk geometries generated by applying a conformal transformation in the boundary carries their own unique defects, which may be associated with symplectic charges or ``Virasoro hair'' \cite{Sheikh-Jabbari:2016unm}. Therefore, this class of Berry phases probes the defects in the dual bulk geometries generated by applying conformal transformations in the boundary. Since such conformal transformations change the state in the field theory, these Berry phases were referred to as state-changing in \cite{deBoer:2021zlm}.

When coupling two such Virasoro coadjoint orbits, the individual holonomies associated with these orbits lead to a spacetime wormhole. We illustrate this explicitly using a model of wormholes considered in \cite{Henneaux:2019sjx}, where a system of two coupled chiral bosons was shown to be mapped to two Virasoro coadjoint orbits. We interpret this map as a ``factorization map'' \cite{Jafferis:2019wkd} to explain how the holonomy of the coupled system interpreted as the signature of a bulk spacetime wormhole can actually be identified with the holonomies of the individual coadjoint orbits. This can therefore be thought of as a consequence of improper diffeomorphisms in the bulk.

The second type of Berry phase, namely the {\bf gauge Berry phase}, results from a different type of bulk diffeomorphism than above. The picture is as follows. The gauge symmetries in the field theory do not yield physically distinguishable states. In the bulk, these gauge transformations correspond to {\it proper diffeomorphims}, which leave the bulk spacetime invariant and realize the Killing symmetry of the dual geometry. {\it Such a diffeomorphism does not yield a Berry phase within a single CFT, but nevertheless yields a Berry phase when two such CFTs are entangled.} Examples of this type of Berry phase are discussed in the context of a gravitational wormhole in \cite{Verlinde:2020upt, Nogueira:2021ngh}.

The third type of Berry phases we discuss are the  {\bf modular Berry phases} \cite{Czech:2017zfq,Czech:2018kvg,Czech:2019vih}. These arise from entanglement between subregions of a single CFT when the subregion in a CFT is deformed continuously such that the deformations form a closed loop and were referred to as shape-changing Berry phases in \cite{deBoer:2021zlm}. Such a setup corresponds to parallel transporting the modular Hamiltonian associated to each subregion around a closed loop. Since the modular Hamiltonian generates time evolution with respect to the abstract modular time, a gauge symmetry is present in each subregion and their complement due to the independent choice of modular time in  each subregion. The Berry curvature was shown to be related to the Crofton form on kinematic space for an interval in a vacuum CFT in \cite{Czech:2019vih}. {\it We derive modular Berry phases for various thermal CFTs and find similar relations to the Crofton form. Furthermore, for an eternal AdS black hole, where we place an interval in each CFT on the boundaries, the Berry curvature exhibits a transition very similar to the transition in the time evolution of the entanglement entropy \cite{ Hartman:2013qma}.} Before the transition, the modular Berry curvature exhibits the same characteristic features of a wormhole in terms of a non-exact symplectic form. 

The paper is organized as follows. In sec. \ref{sec:Virasoro_Berry_phase} we discuss Virasoro Berry phases in the framework of coadjoint orbits and discuss their bulk interpretation. These Berry phases arise from improper diffeomorphisms. We show that coupling two Virasoro coadjoint orbits gives rise to a spacetime wormhole and derive the corresponding non-exact symplectic form, which may be interpreted as the Berry curvature. Furthermore, we discuss how factorization of such coupled systems is achieved. In sec. \ref{sec:gauge_Berry_phase}, we discuss gauge Berry phases, which originate from proper diffeomorphisms. We demonstrate that such diffeomorphisms lead to Berry phases in entangled CFTs, corresponding to a spacetime wormhole in the bulk. In sec. \ref{sec:modular_Berry_phase} we derive modular Berry phases for a single interval in a thermal CFT on the cylinder and on the torus, dual to an AdS black hole and black string, respectively. We also consider two intervals in entangled CFTs dual to an eternal AdS black hole and observe a transition similar to that observed in the time evolution of the entanglement entropy in \cite{Hartman:2013qma}. Before the transition, we recover a non-exact symplectic form associated to a wormhole.  

\section{Virasoro Berry phase}
\label{sec:Virasoro_Berry_phase}
We start by describing the Berry phase in a single coadjoint orbit and show how coupling two such orbits gives rise to a spacetime wormhole in the gravity dual. Using a model of two coupled chiral bosons, derived in \cite{Henneaux:2019sjx} as the asymptotic dynamics of the BTZ black hole, we demonstrate how an effective factorization of the Hilbert space can be realized in terms of reduction of the full symmetry to its diagonal subgroup.

\subsection{Berry phase as the holonomy of a coadjoint orbit}
\label{subsec:Virasoro_coadjoint orbit}
We begin by introducing Virasoro Berry phases as considered in \cite{Oblak:2017ect} in the framework of coadjoint orbits \cite{Alekseev:1988ce, Alekseev:1990mp,Witten:1987ty}. We then discuss their interpretation in the dual AdS geometries, employing the classification of bulk geometries in terms of Virasoro coadjoint orbits presented in \cite{Compere:2015knw, Sheikh-Jabbari:2016unm}.

The concept of states that acquire a phase when parallel transported around a closed loop in parameter space can be generalized to symmetry groups. Given a symmetry group $G$ with group elements $g\in G$ and a highest-weight state $\ket{\phi}$, there is a subgroup of transformations in $G$ which leave invariant the state $\ket{\phi}$. This constitutes a gauge redundancy, and the subgroup $H\subset G$ formed by these gauge transformations is called the stabilizer group. When the state $\ket{\phi}$ is parallel transported around a closed loop in the group manifold by continuously applying group transformations  $\ket{\phi(t)}=U_{g(t)}\ket{\phi}$, the state will generally not return to its initial form $\ket{\phi}$, but differ by a phase due to the gauge redundancy introduced by the stabilizer group $H$, $\ket{\phi}\rightarrow e^{i\alpha}\ket{\phi}$. This gives rise to a Berry phase. The Berry connection relates different points in the tangent space of the group manifold and is therefore given in terms of the Maurer-Cartan form $\theta$ of the symmetry group,
\begin{equation}
A=	i\bra{\phi(t)}d\ket{\phi(t)}=i\bra{\phi}\mathfrak{u}(\theta)\ket{{\phi}},
\label{eq:Berry_connection_symmetrygroup}
\end{equation} 
where $d$ denotes the exterior derivative on the group manifold and we have used that $U_g^\dagger dU_g= \mathfrak{u}(\theta)$.
This concept can be applied straightforwardly to any symmetry group, including the Virasoro group \cite{Oblak:2017ect}. For the latter, the symmetry group is the centrally extended group of the diffeomorphisms of the unit circle $\mathrm{Diff}(S^1)$ with group elements $(f(\phi),\alpha)$, where $f(\phi)$ are diffeomorphisms of $S^1$ and $\alpha\in \mathbb{R}$. Highest-weight states are denoted $\ket{h}$ for conformal weights $h>0$ and $\ket{0}$ for the vacuum with respective stabilizer subgroups $U(1)$ and $SL(2,R)$ that generate the gauge redundancy giving rise to the Berry phase. There is a subtlety in defining the Virasoro Berry connection. The formula \eqref{eq:Berry_connection_symmetrygroup} cannot be straightforwardly applied but needs to be supplemented by an additional term since the group is centrally extended. Group elements are pairs, and while $\alpha$ does not contribute to the Berry phase itself, the Maurer-Cartan form picks up a central extension as well, $(\theta,m_\theta)$. Therefore, the connection reads
\begin{equation}
A=i\bra{h}\mathfrak{u}(\theta)+c\mathfrak{u}(m_\theta)\ket{h},
\label{eq:Berry_connection_Vir}
\end{equation}
where $c$ denotes the central charge of the CFT. 
Since a continuous group path of conformal transformations $U_{f(t)}$ generate states within the same Verma module, Virasoro Berry phases may be thought of as probing the ``geometry'' of the Verma module. Verma modules may be obtained by quantizing coadjoint orbits of the Virasoro group \cite{Alekseev:1988ce, Alekseev:1990mp,Witten:1987ty}, which are given by the manifold $\mathrm{Diff}(S^1)/H$ with $H$ the stabilizer subgroup $ U(1)$ or $SL(2,\mathbb{R})$. For a self-contained presentation, we briefly introduce the orbit construction. Coadjoint orbits of the Virasoro group are obtained by taking an element of the dual Lie algebra $b_0\in \mathfrak{g}^*$ and applying all possible group transformations to this element by a coadjoint transformation
\begin{equation}
\mathcal{O}_{b_0}=\{b=\mathrm{Ad}^*_{(f,\alpha)}b_0|f\in \mathrm{Diff}(S^1)\},
\end{equation}   
where 
\begin{equation}
\mathrm{Ad}^*_{(f,\alpha)}b_0=f'^2b_0-\frac{c}{24\pi}\{f,\phi\}
\label{eq:coadjoint_orbit_transformation}
\end{equation}
and $\{\cdot,\cdot\}$ is the Schwarzian derivative. Depending on the particular orbit label $b_0$, there are transformations which leave the constant orbit representative $b_0$ invariant. These elements form the stabilizer subgroup $H$. Therefore, coadjoint orbits are manifolds $\mathrm{Diff}(S^1)/H$. Furthermore, the coadjoint orbits of the Virasoro group form symplectic manifolds on which we may define a symplectic form and potential. The symplectic form is given by
\begin{equation}
\omega =-d \langle (b,c),(\theta,m_\theta) \rangle ,
\end{equation} 
where $\langle \cdot, \cdot \rangle $ is the bracket between the Lie algebra and the dual Lie algebra. The symplectic potential is then given by
\begin{equation}
\alpha =-\langle (b,c),(\theta,m_\theta) \rangle.
\end{equation}
When evaluated explicitly for the Virasoro group, $\alpha$ is just the Berry connection  \eqref{eq:Berry_connection_Vir}. Therefore, Virasoro Berry phases probe the geometry of the particular coadjoint orbit. In particular, the Berry connection \eqref{eq:Berry_connection_Vir} is the symplectic potential on the coadjoint orbit $\alpha$ and the curvature $F=dA$ the symplectic form $\omega$. Such Berry phases have also been found for certain choices of cost functions in circuit complexity \cite{Caputa:2018kdj, Erdmenger:2020sup}.

For holographic CFTs, it is interesting to ask which feature of the bulk Virasoro Berry phases probe. To this end, note that the energy-momentum tensor $T(\phi)$ of the CFT is an element of the dual Lie algebra $\mathfrak{g}^*$, and \eqref{eq:coadjoint_orbit_transformation} is nothing but the expectation value of the transformed energy-momentum tensor under conformal transformations $\tilde{\phi}=f(\phi)$,
\begin{equation}
2\pi b=2\pi \mathrm{Ad}^*_{(f,\alpha)}b_0=\bra{h}\tilde{T}(\phi)\ket{h}=f'^2\bra{h}T(\tilde{\phi})\ket{h}-\frac{c}{12}\{f,\phi\},
\end{equation}
where we have identified 
\begin{equation}
b_0=\frac{1}{2\pi}\bra{h} T(\tilde{\phi})\ket{h}=\frac{1}{2\pi}\left(h-\frac{c}{24}\right).
\end{equation}
The gauge transformations on the orbit leaving invariant the orbit representative $b_0$ may then be interpreted as those transformations which leave invariant the expectation value of the energy-momentum tensor. Depending on whether the expectation value is taken in the state $\ket{h}$ with $h>0$ or the vacuum $\ket{0}$, these transformations form precisely the stabilizer groups $U(1)$ or $SL(2,\mathbb{R})$, respectively.

Having established the relation between $b$ and the transformed energy-momentum tensor, we may proceed to identify dual bulk geometries. In AdS$_3$/CFT$_2$, the dual bulk geometries may be constructed from the Fefferman-Graham expansion \cite{AST_1985__S131__95_0, deHaro:2000vlm}, which takes as input only the boundary metric, set to a flat metric, and the expectation value of the energy-momentum tensor, given by the coadjoint transformation of the constant orbit representative $b_0$. Since the energy-momentum tensor does not transform under $SL(2,\mathbb{R})$ or $U(1)$ transformations, the gauge redundancies must also be present in the bulk. Furthermore, as the bulk metric is determined in terms of the energy-momentum tensor, all physically distinguishable expectation values of the energy-momentum tensor given in terms of $b=\mathrm{Ad}^*_{(f,\alpha)}b_0$, correspond to different bulk geometries. The dual bulk geometries associated to Virasoro coadjoint orbits are the Banados geometries \cite{Banados:1998gg},
\begin{equation}
d s^{2}=\ell^{2} \frac{d r^{2}}{r^{2}}-\left(r d x^{+}-\ell^{2} \frac{L_{-}\left(x^{-}\right) d x^{-}}{r}\right)\left(r d x^{-}-\ell^{2} \frac{L_{+}\left(x^{+}\right) d x^{+}}{r}\right),
\end{equation}  
where $L(x^\pm)=\frac{6}{c}\bra{h}\tilde{T}\ket{h}=\frac{12\pi}{c}b$ and $x^\pm=t\pm\phi$ \footnote{In the dual gravity description, we employ $T(x^\pm)$ rather than $T(\phi)$. Both are physically equivalent. }. 
The Banados geometries are the set of on-shell solutions to AdS$_3$ gravity with Brown-Henneaux boundary conditions \cite{Brown:1986nw} and and have non-vanishing holographic entanglement entropy \cite{Sheikh-Jabbari:2016znt}. These geometries form an on-shell phase space. To understand, which feature of the bulk the Virasoro Berry phases probe, it is necessary to understand the structure of this phase space. The phase space was analyzed in detail in \cite{Compere:2015knw} and may be classified in terms of conserved charges. Since the Banados geometries form an on-shell phase space, we may define a presymplectic form on this phase space.  

In gravity, a charge $Q_\xi$ for an on-shell metric perturbation $\delta_\xi g_{\mu\nu}$ by a vector field $\xi$ is conserved if and only if the presymplectic form on the on-shell gravitational phase space vanishes  
\begin{equation}
\mathbf{\omega}[\delta_\xi g_{\mu\nu}, \delta g_{\mu\nu},g_{\mu\nu}]=0.
\end{equation}
There are three different conserved charges that arise from this condition: Killing charges, which are generated by proper diffeomorphisms, i.e. $\delta_\xi g_{\mu\nu}=0$. These are global symmetries of the bulk geometry specified by the metric $g_{\mu\nu}$ as these transformations leave the bulk metric invariant. Then, there are asymptotic symmetries, for which the condition $\delta_\xi g_{\mu\nu}=0$ is only satisfied as we approach the asymptotic boundary. For these charges,  $\mathbf{\omega}[\delta_\xi g_{\mu\nu}, \delta g_{\mu\nu},g_{\mu\nu}]=0$ only holds in the asymptotic region. Finally, there are symplectic charges, for which  $\delta_\xi g_{\mu\nu}\neq 0$ but $	\mathbf{\omega}[\delta_\xi g_{\mu\nu}, \delta g_{\mu\nu},g_{\mu\nu}]=0$ everywhere in the bulk spacetime. These charges are generated by improper diffeomorphisms, which -- unlike Killing symmetries -- are not isometries of the bulk and therefore change the bulk geometry but nevertheless give rise to conserved charges.  Note that in order to define the charges, we must specify with respect to which presymplectic form on the phase space these charges are defined, i.e. which presymplectic form vanishes for a given metric perturbation. In principle, there are two choices: The Lee-Wald symplectic form \cite{Lee:1990nz} giving rise to Iyer-Wald surface charges \cite{Iyer:1994ys} or the invariant presymplectic form with conserved Barnich-Brandt charges \cite{Barnich:2001jy,Barnich:2007bf}. It was shown in \cite{Compere:2015knw, Sheikh-Jabbari:2016unm} that for the phase space formed by the Banados geometries, charges may be defined with respect to the invariant presymplectic form. There are two types of metric perturbations for which the invariant presymplectic form vanishes: Killing charges generated by proper diffeomorphisms leaving invariant the metric ($\delta_\xi g_{\mu\nu}= 0$ ) and symplectic charges, called ''Virasoro hair'', generated by improper diffeomorphisms moving us among different Banados geometries on the phase space ($\delta_\xi g_{\mu\nu}\neq 0$ ). In particular, the Killing charges correspond to the gauge redundancy of $\bra{h}\tilde{T}(\phi)\ket{h}$ and hence form the stabilizer subgroup $H$. For instance, the vacuum state $\ket{0}$ has stabilizer group $SL(2,\mathbb{R})$. It is well known that the dual bulk geometry, which is empty AdS, has an $SL(2,\mathbb{R})$ Killing symmetry. Banados geometries arising from different values of  $\bra{h}\tilde{T}(\phi)\ket{h}$ in the Fefferman-Graham expansion may be distinguished by their symplectic charge \cite{Compere:2015knw}, called Virasoro hair,
\begin{equation}
Q_{\chi}[g]=\int_{S} \boldsymbol{Q}_{\chi}[g]=\frac{\ell}{8 \pi G} \int_{0}^{2 \pi} d \phi\left(\epsilon_{+}\left(x^{+}\right) L_{+}\left(x^{+}\right)+\epsilon_{-}\left(x^{-}\right) L_{-}\left(x^{-}\right)\right).
\end{equation}
These charges are defined by integration around non-contractible circles in the bulk. The Virasoro hair may thus be associated to a particular defect in the bulk. Note that these defects also exist for descendants of global AdS, i.e. the dual geometries to descendants of the CFT vacuum. Therefore, from the bulk point of view, picking a closed path $f(t)$ on a particular orbit that gives rise to a Virasoro Berry phase corresponds to moving among different bulk geometries in a class of geometries with identical Killing charge but different Virasoro hair. The Berry phase thus probes the defects giving rise to non-vanishing Virasoro hair. 

As an example consider the orbit $\mathcal{B}_0$, which is labeled by $b_0=\frac{a}{4}=\frac{1}{2\pi}\left(h-\frac{c}{24}\right)$, where $a>0$, and similarly for the right-moving sector. In the bulk, this corresponds to a BTZ black hole. The spacetime is invariant under $U(1)$ transformations, which constitutes a Killing symmetry of the bulk. In the boundary theory, the state is given by $\ket{h}$ with $h> 0$. Hence, the orbit is also invariant under $U(1)$ transformations, $b_0=\mathrm{Ad}^*_{(f,\alpha)}b_0$, where $f\in U(1)$. This illustrates that the stabilizer subgroup of a particular orbit corresponds to a Killing symmetry in the bulk. In the bulk, these symmetries are generated by proper diffeomorphisms. These leave invariant the metric. The points on the orbit generated by applying coadjoint transformations to the reference point $b_0$, $b=\mathrm{Ad}^*_{(f,\alpha)}b_0=\frac{1}{2\pi}\bra{h}\tilde{T}(\phi)\ket{h}$, have the same Killing charge but different symplectic charges in the bulk. Since the appropriate bulk diffeomorphims generated by the symplectic charges map different Banados geometries to one another -- each labelled by a different symplectic charge but identical Killing charge -- these diffeomorphisms are improper diffeomorphism. 

Therefore, we conclude that Virasoro Berry phases picked up by a state when moving around a closed loop (up to improper diffeomorphisms) through the group manifold probe the geometry associated to a particular coadjoint orbit and upon quantization the Verma module. In the bulk spacetime, we move on a closed path among different Banados geometries which have the same Killing charge but different symplectic charges. These symplectic charges arise from defects in the bulk. In the bulk spacetime, the Berry phase thus probes the defects in the bulk spacetime. Note that this notion of Berry phase is unrelated to entanglement. Since Virasoro Berry phases involve transformations of the energy-momentum tensor induced by conformal transformations, such Berry phases are called state-changing in \cite{deBoer:2021zlm}.

\subsection{Interaction, entanglement and Berry phase in general quantum systems}
With our understanding of the appearance of Berry phases in a single Virasoro coadjoint orbit, we are now in a position to analyze how a coupling between two identical orbits gives rise to a non-exact symplectic form in the CFT Hilbert space, signalling the presence of a spacetime wormhole in the gravity dual. Following \cite{Verlinde:2021kgt}, we start with the general identification of a wormhole in terms of a non-exact symplectic form in the partition function of a general quantum mechanical system. In \secref{sec:entangled-CFT} we will use this definition to identify the non-exact symplectic form appearing in a model of two coupled chiral bosons, derived in \cite{Henneaux:2019sjx} as the asymptotic dynamics of the BTZ black hole. Employing a map of this coupled system to the action of two Virasoro coadjoint orbits, we demonstrate how an effective notion of factorization emerges in the CFT Hilbert space. 
\subsubsection{The wormhole partition function}
\label{sec:Verlinde}
Wormholes play an important role in understanding the AdS/CFT correspondence, one of the first insights being the relation between the eternal non-traversable wormhole in AdS spacetime and the TFD state \cite{Maldacena:2001kr}; a recent review on wormholes in holography was written in \cite{Kundu:2021nwp}. However as discussed in \cite{Verlinde:2021kgt}, the notion of a wormhole is not confined to theories of gravity, but extends to generic quantum mechanical systems with a non-exact symplectic form. In order to put our discussion in the following section into proper context, we will first briefly review this notion of wormholes in the following.\\
The thermal partition function of a quantum system is written as a path integral in terms of the Hamiltonian $H$ and the symplectic form $\Omega$ as
\begin{equation}
	Z(\beta)=\int[dX]\exp\left(\int_D\Omega-\oint_{\partial D}Hdt\right),
\end{equation}
where $D$ is a disk with the thermal circle $\partial D$ as boundary\footnote{Note that, since time is $\beta$-periodic, this discussion refers to Euclidean wormholes.}. If the symplectic form is exact, by use of Stokes theorem the integral in the exponent reduces to the usual one dimensional path integral of the quantum system. However if $\Omega$ is not exact, i.e. it cannot be written as $\Omega=\text{d}\alpha$ globally, the thermal partition function corresponds to the functional integral on $D$. For such systems, wormholes can be introduced by replacing the disk $D$ with a general two dimensional manifold $\Sigma$ which has $n$ thermal circles as boundaries. The path integral \cite{Verlinde:2021kgt}
\begin{equation}
	Z(\Sigma)=\int[dX]\exp\left(\int_\Sigma\Omega-\oint_{\partial\Sigma}Hdt\right)\label{eq:PartFunctionVerlinde}
\end{equation}
then includes wormhole contributions. In particular, the path integral for the $n$-fold trumpet geometry $\Sigma_n$ corresponds to $\langle Z(\beta)^n\rangle$. This only factorises for an exact symplectic form $\Omega=d\alpha$, in which case $Z(\Sigma_n)=Z(D)^n$. Examples for such quantum systems yielding a wormhole contribution range from two dimensional CFTs to very simple scenarios such as two coupled harmonic oscillators, as discussed in the appendix of \cite{Verlinde:2021kgt}.

As pointed out, this construction works both for generic quantum systems as well as for systems with a gravitational dual. Therefore, in this work we distinguish the corresponding types of wormholes: if the system under consideration has a gravitational dual, we speak of spacetime wormholes, while in other cases, we refer to it as topological wormhole. In the following sections, we discuss how these features are found in the CFT Hilbert space for a specific model of two coupled bosons.

\subsection{Virasoro Berry phase for two entangled CFTs}
\label{sec:entangled-CFT}

In \secref{sec:Verlinde} we described how a wormhole partition function can be defined for generic quantum systems following \cite{Verlinde:2021kgt}. In this section we will explain how the features of this construction can also be found in holographic scenarios. To do so, we consider the coupled chiral boson model, derived in \cite{Henneaux:2019sjx} as the dual theory to three-dimensional gravity on an annulus topology. After describing this model and its wormhole features in detail, we move on to explain how the Virasoro Berry phase can also be found in this model. We further interpret these results in terms of factorization, and in particular the factorization map.

\subsubsection{An illustrative glance at U(1) Chern-Simons theory on the annulus}
\label{sec:U(1)}
As discussed in \secref{sec:Verlinde}, non-factorization presents us with an intriguing problem inherently related to the presence of wormholes. Non-factorization can also be observed in AdS$_3$/CFT$_2$. This was shown in \cite{Cotler:2018zff,Henneaux:2019sjx,Cotler:2020ugk}. We now show that new insights into this problem are gained by interpreting the results of \cite{Henneaux:2019sjx} in the context of Berry phases that arise in wormhole geometries from non-exact symplectic forms introduced in \cite{Verlinde:2020upt, Verlinde:2021kgt}.\\
A particular simple example, which already exhibits all the features of non-exactness necessary for non-trivial Berry phases is a $U(1)$ Chern-Simons theory on the annulus, which was derived in \cite{Henneaux:2019sjx}. The two boundaries of the annulus represent the two asymptotic boundaries at a fixed time with $U(1)$ Kac-Moody algebras. The interior of the annulus represents a constant timeslice of the bulk spacetime. Due to the topology of an annulus, it is possible to obtain non-trivial holonomies. Therefore, upon deriving the action for the boudary field theories from the bulk Chern-Simons theory, one finds a set of fields that can be devided into two different types: Local fields that live on either of the two boundaries and non-local fields. Of particular importance in the context of Berry phases arising from non-exact symplectic forms are the non-local fields. These non-local fields arise necessarily from the bulk point of view since the holonomy must give the same result irrespective from which boundary it is measured. Therefore, there must exist at least one field that couples the actions on both boundaries. The action derived in \cite{Henneaux:2019sjx} describes two chiral bosons coupled by the holonomy $k_0$ of the annulus, 
\begin{equation}
	\begin{aligned}
		S&=
		\frac{k}{4 \pi} \left(\int d t \left[ \oint d \varphi\left(\partial_{\varphi} \Phi \partial_{t} \Phi\right)-H_{\Phi} \right]
		+ \int d t\left[-\oint d \varphi\left(\partial_{\varphi} \Psi \partial_{t} \Psi\right)-H_{\Psi}\right] \right.\\
		&+\left.2\int d t\left[\oint d \varphi k_{0}\left(\partial_{t} \Phi-\partial_{t} \Psi\right)-H_{0}\right]\right),
	\end{aligned}\label{eq:U1ChiralBosonsAction}
\end{equation}
where
\begin{equation}
	\begin{aligned}
		H_{\Phi}&=\int d \varphi\left(\partial_{\varphi} \Phi\right)^{2},\\
		H_{\Psi}&=\int d \varphi\left(\partial_{\varphi} \Psi\right)^{2},\\
		H_{0}&=2 \pi\left(k_{0}\right)^{2}.
	\end{aligned}\label{eq:U1ChiralBosonsHamiltonians}
\end{equation}
The canonical conjugate to the holonomy is given by \cite{Henneaux:2019sjx}
\begin{equation}
	\Pi_{0}=-\frac{k}{2 \pi} \oint d \varphi(\Phi-\Psi)=-\frac{k}{2 \pi} \oint d \varphi\left(\int_{r_{1}}^{r_{2}} d r A_{r}\right).
	\label{eq:conjugate_holonomy}
\end{equation}
We note that the holonomy $k_0$ appears only via the topology of the bulk space and is intrisically non-local from the point view of the dual CFTs, since it is not associated to just one of these CFTs. 
Moreover, note that $\Pi_0$ is directly related to the radial Wilson line connecting the two boundaries at $r_1$ and $r_2$. Using the relation
\begin{equation}
	\dot{x}^{a}=\left(\omega^{-1}\right)^{a b} \partial_{b} H_{\text{full}},
	\label{eq:symplectic_form}
\end{equation}
where
\begin{equation}
	H_{\text{full}}=\frac{k}{4\pi}\left(H_\Phi+H_\Psi+2H_0\right),
\end{equation}
the symplectic form on the boundary phase space $x=(\Phi,\Psi, \Pi_0, \Pi_\Phi, \Pi_\Psi, k_0)$ follows as
\begin{equation}
	\omega=d\Pi_\Phi\wedge d\Phi+d\Pi_\Psi\wedge d\Psi+dk_0\wedge d\Pi_0.\label{eq:abelian_symplectic form_intermediate}
\end{equation}
Note that this depends on the holonomy $k_0$. This term, which is non-local from the point of view of the dual CFTs, gives rise to a non-exact contribution to the symplectic form 
\eqref{eq:abelian_symplectic form_intermediate}. In particular, the last term is identical to the one found for JT gravity in \cite{Harlow:2018tqv} with the identifications
\begin{equation}
	\begin{aligned}
		\phi_h\, &\widehat{=} \, \sqrt{\phi_b\pi}k_0(t),\\
		\delta \, &\widehat{=} \, \Pi_0(t),\\
		H_{\text{JT}}=\frac{2\phi_h^2}{\phi_b} \, &\widehat{=} \,H_0=2\pi k_0^2.
	\end{aligned}
\end{equation}  
Moreover, upon setting $k_0=0$ the symplectic form reduces to 
\begin{equation}
	\omega=d\Pi_\Phi\wedge d\Phi+d\Pi_\Psi\wedge d\Psi.
	\label{eq:symplectic_form_k0_vanishes}
\end{equation}
A vanishing holonomy means that the topology is no longer an annulus but rather a disk. \eqref{eq:symplectic_form_k0_vanishes} shows that in this case the symplectic form becomes exact again, i.e. all non-local pieces drop out and the boundary factorizes. Therefore, for $k_0=0$ the symmetry is $\frac{\widehat{LG}}{\mathrm{G}}\times \frac{\widehat{LG}}{\mathrm{G}}$ with $\widehat{L_G}$ the centrally extended loop group, i.e.~the Kac-Moody group, whereas for non-vanishing $k_0$ the boundaries are connected by a Wilson line and coupled by the holonomy. The latter case gives rise to the symmetry $\frac{\widehat{LG}\times \widehat{LG}}{\mathrm{G}}$, i.e.~the boundary symmetry no longer factorizes. From this point of view, it is the common quotient that leads to the non-exact symplectic form from which we may obtain non-vanishing Berry phases. This concludes our first example for a Berry phase and its associated non-local variable $k_0$ characterizing a wormhole. 

\subsubsection{Generalizing to $SL(2,R)$}
\label{sec:generalizing}

We now consider a different example and apply the  procedure introduced in the preceding subsection to the action describing the asymptotic dynamics of an $SL(2,R)$ Chern-Simons action after boundary conditions for asymptotic AdS$_3$ have been imposed.
Beginning with an $SL(2,R)\times SL(2,R)$ Chern-Simons theory on the annulus, the authors of \cite{Henneaux:2019sjx} showed that once $AdS_3$ boundary conditions are imposed the action describing the asymptotic dynamics of a BTZ black hole is given by
\begin{equation}
	S\left[k_{0}, \Phi, \Psi\right]=\frac{k}{4 \pi} \int d t d \varphi\left(\frac{1}{2} \partial_{-} \Phi \Phi^{\prime}-\frac{1}{2} \partial_{+} \Psi \Psi^{\prime}+k_{0}\left(\partial_{-} \Phi-\partial_{+} \Psi\right)-k_{0}^{2}\right).
	\label{eq:action:non_Abelian}
\end{equation}
Note that we consider only one of the $SL(2,R)$ symmetries, the other one gives an analogous action.
Similar to the $U(1)$ case the action is again given in terms of two chiral bosons, $\Phi$ is the field on the outer boundary and $\Psi$ on the inner boundary, coupled by a holonomy $k_0$. The conjugate momentum to the holonomy $k_0$ is up to a prefactor identical to  \eqref{eq:conjugate_holonomy} and is given by \cite{Henneaux:2019sjx}
\begin{equation}
	\Pi_{0}=-\frac{k}{4 \pi} \int d \varphi(\Phi-\Psi).
\end{equation}
However, it cannot be related as straightforwardly to the radial Wilson line as in the Abelian case.
Furthermore, the Hamiltonian is given by
\begin{equation}
	H=\frac{k}{4\pi}\left(\frac{1}{2} \int d \varphi \Phi^{\prime 2}+\frac{1}{2} \int d \varphi \Psi^{\prime 2}+2 \pi k_{0}^{2}\right).
\end{equation}
The phase space variables and their conjugate momenta are again $x=(\Phi,\Psi, \Pi_0, \Pi_\Phi, \Pi_\Psi, k_0)$. Employing \eqref{eq:symplectic_form}, we derive the symplectic form as discussed in sec. \ref{sec:U(1)},
\begin{equation}
	\omega =d\Pi_\Phi\wedge d\Phi +d\Pi_\Psi \wedge d\Psi+dk_0\wedge d\Pi_0.
	\label{eq:symplectic form_BTZ}
\end{equation} 
The result is completely analogous to the Abelian case. Furthermore, the interpretation is also the same. Both sides are coupled by the holonomy and are connected by a radial Wilson line. This implies that the symmetry group is enhanced from $\frac{\mathrm{Diff(S^1)}}{S^1}\times\frac{\mathrm{Diff(S^1)}}{S^1}$ for two separate orbits to $\frac{\mathrm{Diff(S^1)}\times \mathrm{Diff(S^1)}}{S^1}$ and the boundary no longer factorizes. If we set $k_0=0$ the actions decouple and the boundary factorizes. This corresponds to a disk topology since the holonomy vanishes. The appearance of non-local terms in \eqref{eq:symplectic form_BTZ} implies that the symplectic form is non-exact and we obtain non-trivial Berry phases.

The action \eqref{eq:action:non_Abelian} can be written as the difference of the actions on the inner and outer boundary ($\Sigma_i$ and $\Sigma_o$, respectively),
\begin{equation}
	\begin{aligned}
		S[k_0,\Phi,\Psi]&=S_o[k_0,\Phi]-S_i[k_0,\Psi]=\\
		&=\frac{k}{4\pi}\int_{\Sigma_o}dtd\varphi\left(\frac{1}{2}\partial_-\Phi(\Phi^\prime+2k_0)-\frac{1}{2}k_0^2\right)\\
		&~~~-\frac{k}{4\pi}\int_{\Sigma_i}dtd\varphi\left(\frac{1}{2}\partial_+\Psi(\Psi^\prime+2k_0)+\frac{1}{2}k_0^2\right).\label{eq:action_split}
	\end{aligned}
\end{equation}
The authors of \cite{Henneaux:2019sjx} show that by employing a different parameterization of the chiral bosons, the corresponding boundary actions $S_o$ and $S_i$ become the geometric actions on coadjoint orbits of the Virasoro group. Explicitly, this is achieved by parameterizing $\Phi$ and $\Psi$ in terms of two functions $f$ and $g$ respectively, and the holonomy $k_0$ in the following way,
\begin{equation}
	\begin{aligned}
		\Phi&=k_0\big(f(t,\varphi)-\varphi\big)-\ln(-k_0f^\prime(t,\varphi))~~~\text{and}\\
		\Psi&=k_0\big(\varphi-g(t,\varphi)\big)-\ln(k_0g^\prime(t,\varphi)).
	\end{aligned}
	\label{eq:map_coadjoint_orbit}
\end{equation}
Inserting this into \eqref{eq:action_split}, up to a boundary term it reduces to the difference of two geometric actions
\begin{equation}
	S[k_0,\Phi,\Psi]=S_{\text{geo}}^-[f,b_0]-S_{\text{geo}}^+[g,b_0],\label{eq:action_geometric}
\end{equation}
where, with $h\in\{f,g\}$,
\begin{equation}
	S_{\text{geo}}^\pm[h,b_0]=\int d t d \sigma \left(b_0h^\prime\partial_\pm h+\frac{k}{8\pi}\frac{h^{\prime\prime}\partial_\pm h^\prime}{(h^\prime)^2}\right).
	\label{eq:geometric_action}
\end{equation}
The orbit label $b_0$ is related to the holonomy by 
\begin{equation}
	b_{0}=\frac{k}{8 \pi} k_{0}(t)^{2},\label{eq:orbit_label}
\end{equation}
where the equation of motion $\dot{k}_0(t)=0$ enforces $k_0=\mathrm{const}$. This was shown in \cite{Henneaux:2019sjx}  and~\cite{Cotler:2020ugk}. 

We observe that the asymptotic dynamics of the BTZ black hole are related to Virasoro Berry phases. Note that since the holonomy $k_{0}(t)$ is identical on both boundaries, the coadjoint orbits on both boundaries are coupled by the same holonomy $k_0$. Therefore, in \eqref{eq:action_geometric} the same orbit label $b_0$ appears in both terms. This observation is crucial in understanding the factorization properties of the inner and outer boundaries $\Sigma_i$ and $\Sigma_o$. Each geometric action $S_{\text{geo}}^i$ can be understood as the action on coadjoint orbits $\frac{\mathrm{Diff(S^1)}}{\mathrm{S^1}}$. Naively, a linear combination of two geometric actions therefore corresponds to $\frac{\mathrm{Diff(S^1)}}{\mathrm{S^1}}\times\frac{\mathrm{Diff(S^1)}}{\mathrm{S^1}}$. This is indeed true for general orbits $b$ and $b^\prime$. However, as long as $b\neq b^\prime$, the two geometric actions do not follow from the chiral boson action \eqref{eq:action_split} since the holonomies associated to $b$ and $b^\prime$ would differ. The relation of \eqref{eq:action_geometric} to \eqref{eq:action_split} can only be made when the coadjoint orbits coincide. In the chiral boson model, this is ensured by the constraint that the holonomy $k_0$ has to be identical when measured from each boundary. In this case, the symmetry is enhanced to $\frac{\mathrm{Diff(S^1)}\times\mathrm{Diff(S^1)}}{\mathrm{S^1}}$ (remember the discussion below \eqref{eq:symplectic form_BTZ}). In essence, this means that although \eqref{eq:action_geometric} appears factorized, it corresponds to a non-factorized theory, which is ensured by measuring the same holonomy $k_0$ from both boundaries. The theory only factorizes when the holonomy is set to zero, $k_0=0$.

This interpretation can also be associated with the field theoretical generalization of the partition function discussed in \secref{sec:Verlinde}. In the appendix of \cite{Verlinde:2021kgt}, two-dimensional CFTs are discussed. The corresponding wormhole partition function is found not to factorize when the geometry $\Sigma$ for the path integral is taken to be the punctured disk. A punctured disk can be understood analogously to the annulus topology discussed in the chiral boson model above. Paths around the puncture cannot be contracted to a single point, i.e. there is a non-vanishing holonomy around the puncture. This holonomy is related to the minimal length $l_h$ of a geodesic surrounding the puncture. In fact, the holonomy given in \cite{Verlinde:2021kgt} is the same found in \cite{Henneaux:2019sjx}, provided that the geodesic length $l_h$ is related to the holonomy $k_0$ by $l_h=2\pi k_0$\footnote{Note that \cite{Verlinde:2021kgt} is done for the torus. Therefore, to compare with \cite{Henneaux:2019sjx}, the corresponding formulas of \cite{Verlinde:2021kgt} have to be modified by $h\to h-\frac{c}{24}$.}. The theory does not factorize as long as there is a puncture in the disk, that is $l_h\neq0$. If the puncture is removed, $l_h=0$, and so by the relation $l_h=2\pi k_0$ the same conclusion as above is achieved: factorization is achieved by $l_h=0=k_0$.\\
\begin{figure}
    \begin{center}
    \includegraphics[width=0.9\linewidth]{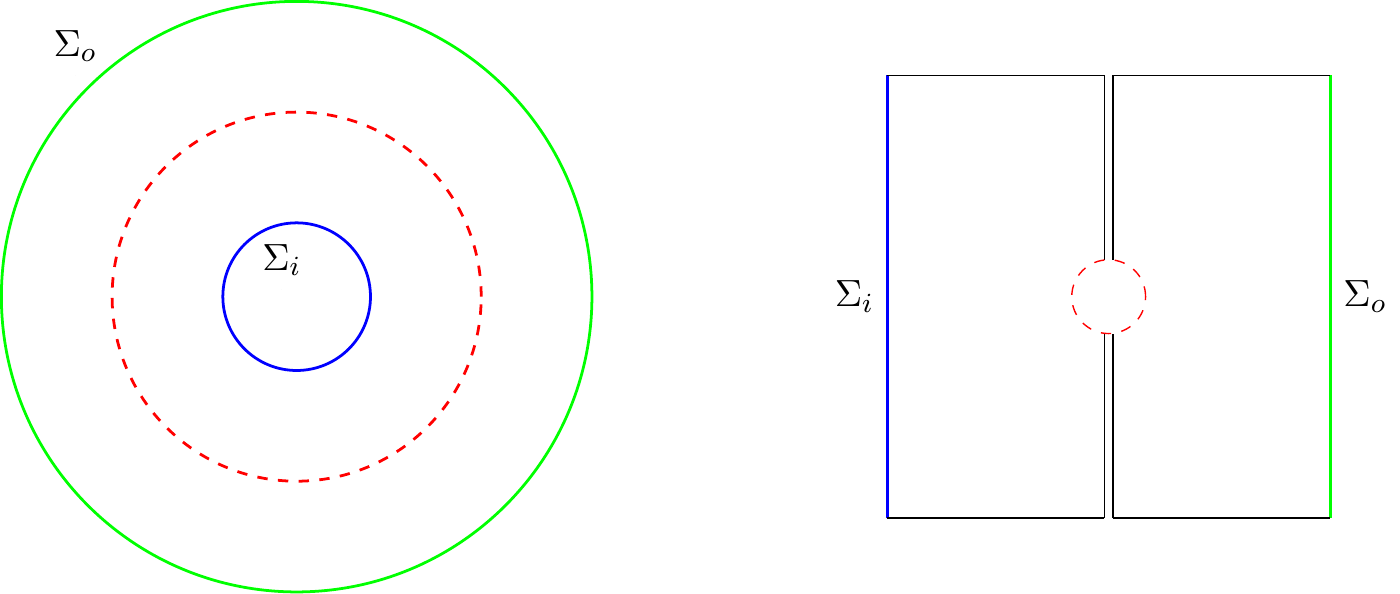}
    \end{center}
    \caption{L.h.s.: fixed time slice of the annulus geometry with the inner and outer boundaries $\Sigma_i$ and $\Sigma_o$, respectively. Along the dashed line, exemplary for a non-contractable circle in $\varphi$ direction, the holonomy $k_0$ is accumulated. The chiral bosons $\Phi$ and $\Psi$ are defined on the boundaries with action given in \eqref{eq:action_split}, or after employing \eqref{eq:map_coadjoint_orbit}, the geometric action in \eqref{eq:geometric_action}. R.h.s.: to interpret the annulus setup in terms of the factorization map, the inner and outer boundaries represent the left and right boundaries. The holonomy, represented again by the red dashed line, is understood as inserting a defect operator which defines the factorization map.}
    \label{fig:Defect_Factorisation}
\end{figure}
We note here that the above described phenomena of diagonalization have an interesting relation with the factorization map. Such a map is used in quantum field theories or quantum gravity to factorize the full Hilbert space. This enables to define quantities like entanglement entropy between the two smaller Hilbert spaces. For JT-gravity, this map is extensively studied in \cite{Jafferis:2019wkd}. It is found that the factorization map in two-dimensional gravity cannot be stated in terms of a local boundary condition for the path integral, but is related to the Euler character via the Gauss-Bonnet theorem. While the study was performed for JT gravity, the general features of a non-local boundary condition are expected to hold for gravity in higher dimensions as well. We illustrate this construction in the right panel of \figref{fig:Defect_Factorisation}, where the red dashed line is understood as inserting a defect operator which serves as non-local boundary condition to factorize the Hilbert spaces of the theories on the boundaries $\Sigma_i$ and $\Sigma_o$. The partition function is then defined by including this defect operator as
\begin{align}
    Z=\tr(De^{-\beta H}),
\end{align}
where $D$ is the defect operator, $\beta$ the inverse temperature and $H$ the Hamiltonian. For JT gravity, the defect operator is related to a winding number constraint \cite{Jafferis:2019wkd}. Analogously, in our discussion above, we saw that only by including the non-local content of the theory, i.e. the holonomy $k_0$, the action \eqref{eq:action:non_Abelian} can be brought into a factorized structure in terms of the actions on the inner and outer boundaries \eqref{eq:action_split}. Therefore we interpret the holonomy contribution in \eqref{eq:action:non_Abelian} as resulting from the defect operator, which in this case would look like
\begin{align}
    D=e^{-\frac{k\beta}{2}k_0^2}.
\end{align}
We hope to come back with a detailed investigation on this in an upcoming work soon.

\section{Gauge Berry phase}
\label{sec:gauge_Berry_phase}
So far we discussed Berry phases originating from improper diffeomorphisms in the bulk. The holonomies in this case are associated with the Virasoro hairs in each of the orbits. The proper diffeomorphisms on the other hand do not produce any holonomy or Berry phase in a single orbit. Rather, these diffeomorphisms are identified as gauge symmetries in the boundary spacetime. As discussed in sec. \ref{subsec:Virasoro_coadjoint orbit}, such diffeomorphisms are characterized by invariance of the metric $\delta_\xi g_{\mu\nu}=0$ and yield Killing charges. However, in a spacetime with a horizon, namely a black hole, there is no global time-like Killing vector and hence instead of a Killing symmetry, one can at most ensure an asymptotic symmetry, namely $\delta_\xi g_{\mu\nu}=0$ locally near the boundary only. This subtle difference can be demonstrated using the example of an eternal black hole. In an eternal AdS black hole, although one can define boundary times, $t_L$ and $t_R$ at the boundary, and can analytically continue at most in the near boundary region, due to the presence of the horizon, it is impossible to define a global bulk time for the whole spacetime. Another way of stating this is, in presence of the horizon, there is no origin of time and accordingly, no time reversal symmetry in the bulk spacetime. The consequences of these asymptotic symmetries, nevertheless, remain the same - they correspond to gauge charges at the respective boundaries of the eternal black holes. Taking into account both boundaries, it is possible to define a Berry phase associated to the proper diffeomorphisms as discussed in \cite{Nogueira:2021ngh}.

This interpretation follows from the analysis of asymptotic boundary conditions in an AdS bulk spacetime. For theories in asymptotically AdS background, we have to fix asymptotic boundary conditions for the dynamical fields to have a well-defined variational principle. The variation of a gravitational action on a manifold $\Cal{M}$ with AdS$_{d+1}$ background has a boundary term
\begin{align}
	\delta S\supset\int_{\partial\Cal{M}}\d[d]{x}\sqrt{\gamma}\,T_{ij}\delta\gamma^{ij},
\end{align}
where $\gamma$ is the metric induced on the boundary and $T_{ij}$ is identified with the boundary energy-momentum tensor. To set this boundary term to zero, typically in AdS/CFT correspondence one imposes Dirichlet boundary conditions on the metric,
\begin{align}
	\gamma^{ij}=\Cal{C}^{ij}.\label{eq:DirichletBdryCondMetric}
\end{align}

Such boundary conditions are not preserved under any arbitrary diffeomorphism. The subset of all diffeomorphisms which do preserve \eqref{eq:DirichletBdryCondMetric} is called the asymptotic symmetry group $\Cal{G}_{\text{asy}}$. A diffeomorphism in $\Cal{G}_{\text{asy}}$ changes the boundary metric at most by further constants $c^{ij}$, such that the variation of $\gamma$ still vanishes. For a Killing vector $\xi$ to belong to $\Cal{G}_{\text{asy}}$ it has to be an isometry of the boundary metric,
\begin{align}
	\gamma_\mu^{~\alpha}\gamma_\nu^{~\beta}\nabla_{(\alpha}\xi_{\beta)}=0.
\end{align}
This follows directly from the invariance of the metric $\delta_\xi g_{\mu\nu}=0$ near the boundary.

\begin{figure}[b]
    \begin{center}
	\includegraphics{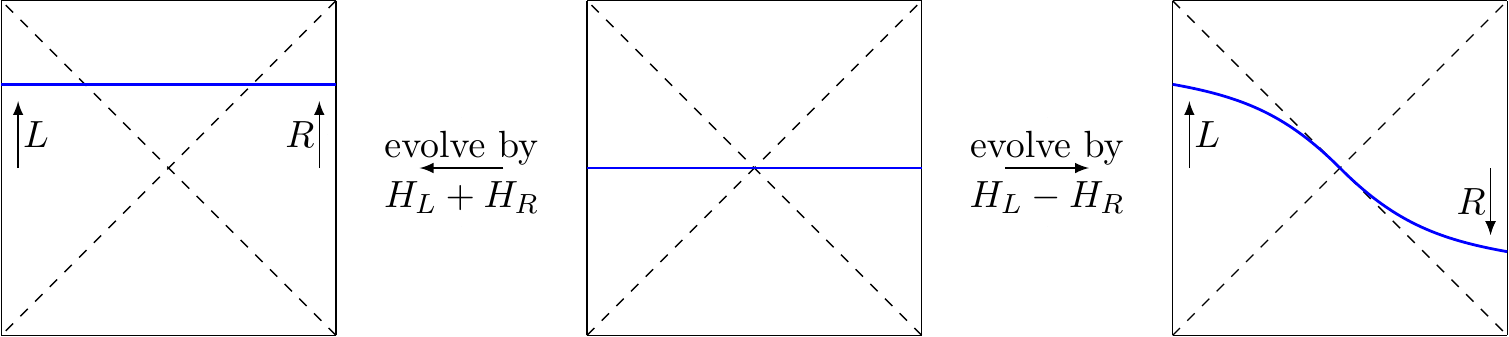}
	\end{center}
	\vspace{-0.5cm}
	\caption{The figure in the left panel shows the holographic dual to the TFD state time-evolved by $H_L + H_R$ while the figure in the right panel shows the time-evolution by $H_L - H_R$. The figure in the middle shows the holographic dual of the original TFD state.}
	\label{fig:time-shiftedTFD}
\end{figure}

\begin{figure}[t]
	\vspace{-0.4cm}
	\begin{center}
		\includegraphics[width=0.45\linewidth]{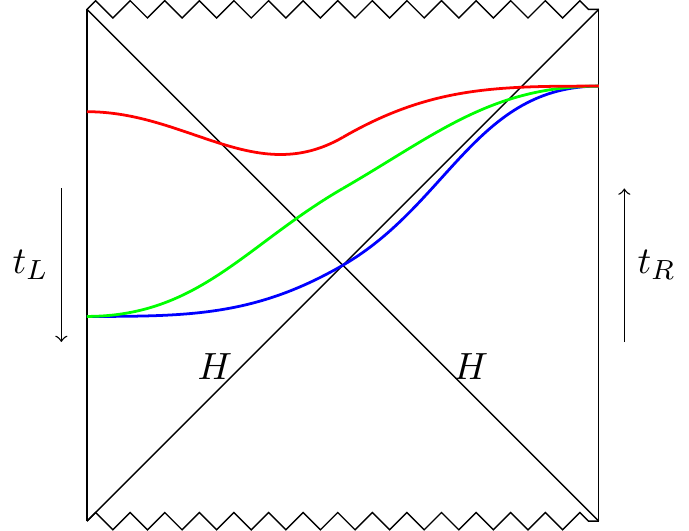}
	\end{center}
	\vspace{-0.5cm}
	\caption{Kruskal diagram of an eternal AdS black hole, jagged lines representing the singularities and $H$ indicating the horizon. The blue and green lines describe the same point in phase space. These two are related by a trivial bulk diffeomorphism which does not induce a time shift at the boundary. The red line is a different point in phase space, corresponding to a different equivalence class of bulk diffeomorphisms.}
	\vspace{-0.2cm}
	\label{fig:AsymptSymmetries}
\end{figure}

As an example, consider the setup of the eternal black hole in AdS spacetime which is conjectured to be dual to a particular pure state, namely, the TFD state \cite{Maldacena:2001kr}
\begin{align}
	\ket{\text{TFD}}=\frac{1}{\sqrt{Z}}\sum_n e^{-\beta\frac{E_n}{2}}\ket{n}_L\ket{n}_R^\ast,\label{eq:TFD-original}
\end{align}
where $Z$ is the canonical partition function, $\beta$ is the inverse temperature and $E_n$ are the sums of the energy eigenvalues, $E_n=(E_n^{(L)}+E_n^{(R)})/2$. $E_n^{L} = E_n^{R}$ are the energy eigenvalues corresponding to the left and right energy eigenstates $\ket{n}_{L/R}^{(\ast)}$. Since $\ket{\text{TFD}}$ is an entangled state of two identical CFT's with isomorphic phase space and identical spectrum, this enjoys a symmetry,
\begin{figure}[t]
    \begin{center}
	\includegraphics[width=0.9\linewidth]{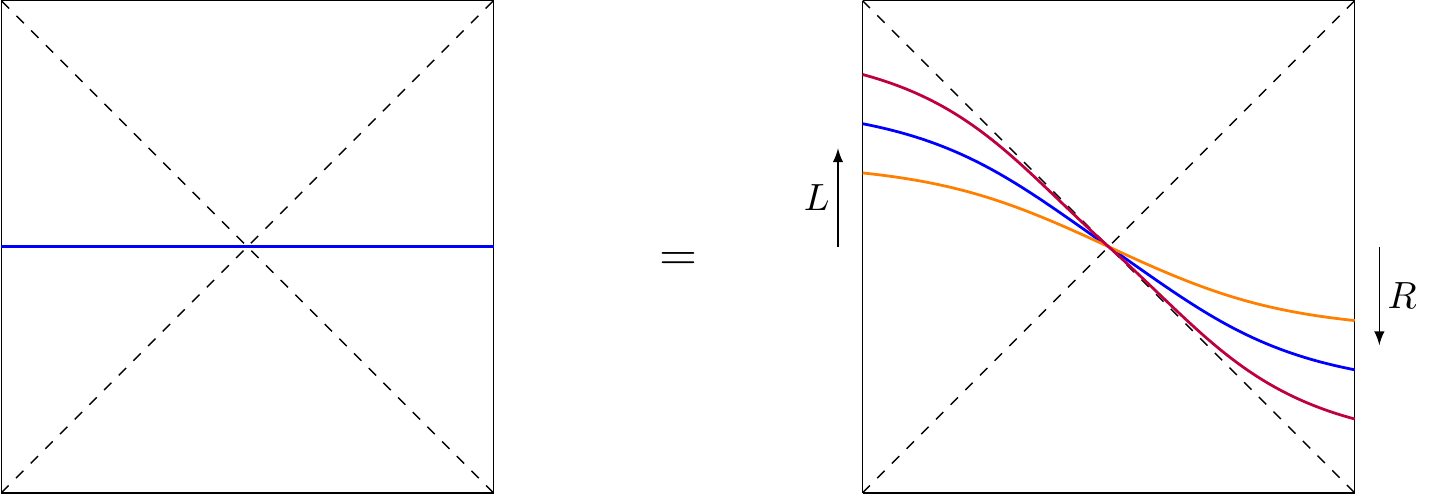}
	\end{center}
	\vspace{-0.5cm}
	\caption{$H_L - H_R$ is the symmetry of the TFD state.}
	\label{fig:HLminusHR}
\end{figure}
\begin{align}
e^{-i (H_L - H_R) \delta} \ket{\text{TFD}} = \ket{\text{TFD}},
\end{align}
which amounts to shifting the left and right boundary times by an equal amount $\delta$ but in opposite directions. This is depicted in \figref{fig:HLminusHR} where all the bulk diffeomorphisms shown on the right correspond to an asymptotic symmetry which approaches a vanishing constant and therefore acts trivially on the phase space. For simplicity, we reduce the discussion to two spacetime dimensions, that is AdS$_2$\footnote{It is possible to construct a TFD state in JT gravity. In this case, apart from the metric, there is also the dilaton field on which appropriate boundary conditions have to be imposed. This does however not interfere with what we discuss in the following.}. The asymptotic boundaries are one-dimensional, each consisting only of the respective time coordinate. The asymptotic symmetries are given by time translations only.
In terms of the additional constants $c^{ij}$ mentioned above, these are $\xi\in\Cal{G}_{\text{asy}}$ which lead to $c^{ij}=0$, which will be called trivial diffeomorphisms in what follows. The $c^{ij}\neq0$ corresponds to a change in the boundary phase space, proportional to these constants. One interesting example of this category is the time-shifted TFD state depicted in the left panel of \figref{fig:time-shiftedTFD} which is obtained by the time evolution of the TFD state with $H_L + H_R$,
\begin{align}
\ket{\text{TFD}_\alpha} = e^{-i (H_L + H_R) \delta}\ket{\text{TFD}},
\label{eq:time-evolution}
\end{align}
where $\ket{\text{TFD}_\alpha}$ can be explicitly written as an entangled state of the left and right energy eigenstates 
\begin{align}
	\ket{\text{TFD}_\alpha}=\frac{1}{\sqrt{Z}}\sum_ne^{i\alpha_n}e^{-\beta\frac{E_n}{2}}\ket{n}_L\ket{n}_R^\ast,\label{eq:TFD-generalized}
\end{align}
with $\alpha_n$ identified as $\alpha_n = - 2 E_n \delta$.
It is worth mentioning here that \eqref{eq:time-evolution} actually defines an equivalence class of diffeomorphisms where the members are related to each other by trivial diffeomorphisms. As a visualisation, let us consider a TFD state with a particular value of $\delta$, for instance the blue line in \figref{fig:AsymptSymmetries} connecting two points on opposite boundaries. Trivial diffeomorphisms connect the same points on the boundaries by a different trajectory through the bulk (blue and green line in \figref{fig:AsymptSymmetries}). Therefore trivial diffeomorphisms will not lead to a change in boundary phase space. Rather, they define the aforementioned equivalence class of states which differ in the bulk but are equivalent from the boundary perspective. To uniquely define the TFD state from the boundary perspective, we therefore mod out $\Cal{G}_{\text{asy}}$ by the trivial diffeomorphisms.

Left over are the non-trivial diffeomorphisms which lead to a change in boundary phase space. Those connect different states in the boundary, such as the red and the blue lines in \figref{fig:AsymptSymmetries}. If one were to also mod out by those, all states would be equivalent; this would reduce the asymptotic phase to just containing the groundstate. We are not interested in this case and keep only the non-trivial asymptotic symmetries.

These time-evolved TFD states have a nice interpretation in the light of ER = EPR \cite{Papadodimas:2015jra, Verlinde:2020upt, Nogueira:2021ngh}. While the phase-shifted TFD states \eqref{eq:time-evolution} correspond to the same bulk metric, the variable $\delta$ provides an additional information, namely, how the boundaries are glued to the bulk geometry. The variable $\delta$ can be identified as the temporal shift $t_L = 2\delta - t_R$ at the horizon accounting for the non-existence of a global time-like Killing vector \cite{Verlinde:2020upt}. This can be treated as the bulk phase space variable and leads to a non vanishing Berry phase
with the Berry connection given by \cite{Nogueira:2021ngh}
\begin{align}
	A_\delta=i\bra{\text{TFD}_\alpha}\partial_\delta\ket{\text{TFD}_\alpha}=\frac{2}{Z}\sum_nE_ne^{-\beta E_n}.\label{eq:BerryConnectionTFD}
\end{align}

This non-zero connection and accordingly, the non-zero Berry phase can in turn be thought of as an observational signature of this wormhole. The horizon is responsible for the non-exactness of the symplectic form which can be explicitly demonstrated using JT gravity in AdS$_2$ \cite{Nogueira:2021ngh}. In fact, in the case of JT gravity, the holonomy resulting in the non-trivial Berry phase can be associated with the winding number on a circle. While the details of this interesting construction and an exact parallel of this with simple quantum mechanical systems can be found in \cite{Nogueira:2021ngh}, for the purpose of the current work, we would like to emphasize that the origin of this class of Berry phases, which we refer to as {\it gauge Berry phases} in our categorization, are very different from the Virasoro Berry phases discussed in the previous section. While the Virasoro Berry phases in entangled systems arise from improper bulk diffeomorphisms in each of the entangled holographic CFTs, the gauge Berry phases result from asymptotic symmetries which are proper bulk diffeomorphisms. Nevertheless, although the origin and the symmetries from which these two kinds of Berry phases result are different, we note that their interpretation in terms of factorization is very similar. We will explain this in more detail when we conclude in \secref{sec:summary_and_conclusion}.

\section{Modular Berry phase}
\label{sec:modular_Berry_phase}
We now move on to consider a third type of Berry phase that arises from entanglement in a single CFT by splitting the CFT into two subregions. These Berry phases, called modular Berry phases, were defined in \cite{Czech:2017zfq, Czech:2018kvg, Czech:2019vih}. Note that these Berry phases differ from those discussed in the previous sections by the transformations that are applied to obtain them. In sec. \ref{sec:Virasoro_Berry_phase}, we considered Berry phases that arise from transformations that change the energy-momentum tensor in the CFT and are thus called state-changing. In sec. \ref{sec:gauge_Berry_phase}, we discussed Berry phases arising from asymptotic gauge transformations. Here, we consider Berry phases that arise from transformations which deform a subregion in a single CFT. We then generalize the modular Berry phase for a single CFT to two entangled CFTs, dual to an eternal AdS black hole, by considering two disjoint intervals, one in each CFT. We show that there is a transition in the Berry curvature induced by a transition in minimal RT surfaces observed in \cite{Hartman:2013qma}. Before the transition occurs, the RT geodesic stretches through the wormhole and connects both boundaries. We show that this gives rise to the Berry phase indicating the presence of a wormhole as in the previous sections, even though here the Berry phase originates from shape deformations of intervals in the subregions and thus of the RT surface. Therefore, Berry phases related to wormholes in the bulk may be obtained from very different types of Berry phases.
  
In sec. \ref{subsec:modular_berry_phase}, we give an introduction into modular Berry phases based on \cite{Czech:2017zfq, Czech:2019vih}. In sec. \ref{sec:Berry curvature thermal} we apply the formalism of \cite{Czech:2019vih} to thermal CFTs, which exhibit a far richer structure than those of the vacuum since RT geodesics bounding the bulk subregion associated to a particular boundary interval exhibit transitions in some thermal CFTs. We calculate the modular Berry curvature for a CFT in a thermal state in the large interval limit in sec. \ref{subsec:thermal}, an interval in a CFT at finite temperature on the cylinder dual to a BTZ black string in sec. \ref{subsec:black_string}, two disjoint intervals in a CFT at finite temperature on the cylinder dual to a two-sided time-evolved black hole in sec. \ref{subsec:2-intervals}, and finally for a BTZ black hole dual to a CFT on the torus in sec. \ref{subsec:thermal_CFT_1_sided}. We demonstrate that the modular Berry curvature exhibits analogous transitions as the entanglement entropy for BTZ black holes and the time-evolved eternal black hole. For the entanglement entropy, these are transitions in the RT geodesic connecting the interval endpoints, whereas for the modular Berry curvature, the transitions correspond to changes in Wilson line configurations. In particular, for the time-evolved eternal AdS black hole, we observe a change in the type of Berry phase associated to the transition in RT geodesic configurations due to the presence of a wormhole in the bulk.  

\subsection{Modular parallel transport}
\label{subsec:modular_berry_phase}
Here, we review relevant aspects of modular Hamiltonians \cite{Jafferis:2015del, Jafferis:2014lza, Bisognano:1975ih, Bisognano:1976za, Cardy:2016fqc}, modular Berry phases \cite{Czech:2017zfq, Czech:2018kvg, Czech:2019vih}, and kinematic space \cite{Czech:2015qta, Asplund:2016koz} that we will use in the next section.

\paragraph{Modular Hamiltonian} 
Given a constant time slice in a 2d CFT and a global state with density matrix $\rho$, we pick a subregion $A$ in the CFT by selecting an interval $[u,v]$. The complement of $A$ is denoted by $\bar{A}$.  The reduced density matrix $\rho_A$ associated to subregion $A$ is obtained by tracing the density matrix $\rho$ associated to the global state over the complement $\bar{A}$, $\rho_A=\tr_{\bar{A}} [\rho]$. Consequently, given some global state, for every choice of subregion $A$ in the CFT, we obtain a reduced density matrix $\rho_A$. Once we have specified the density matrix $\rho_A$, we may define the modular Hamiltonian associated to the particular subregion $A$ as
\begin{equation}
K=- \log \rho_A.
\end{equation}
In general, this modular Hamiltonian is a complicated non-local operator. There exist only a handful of cases in which the modular Hamiltonian is known explicitly. Perhaps the most prominent example is the modular Hamiltonian associated to the half-space $x>|t|$ in a relativistic QFT in the vacuum. This is simply a vacuum QFT in Rindler space. It was shown in \cite{Bisognano:1975ih, Bisognano:1976za} that in this case, the modular Hamiltonian is given in terms of the Rindler boost, 
\begin{equation}
K=2\pi K_{\text{boost}}=2 \pi \int_{x>0} d^{d-1} x\, x T_{tt}(x).
\end{equation}  
Therefore, in any 2d CFT that can be mapped back to the Rindler case, the modular Hamiltonian takes a simple local form
\begin{equation}
K=\int d x f(x) T_{tt},
\label{eq:mod_Ham_local}
\end{equation}
where $f(x)$ is a function related to the components of the boost vector. The CFTs and shape of the intervals for which such a mapping is possible are summarized in \cite{Cardy:2016fqc} and amount to only a handful of cases. Fortunately, the modular Hamiltonians of holographic CFTs exhibit a special feature that sets them apart from their non-holographic counterparts. The holographic dictionary implies that the CFT modular Hamiltonian is related to the entanglement entropy to leading order in $G_N$ \cite{Jafferis:2015del, Jafferis:2014lza},
\begin{equation}
K_{\mathrm{bdy}}=\frac{\text { Area }}{4 G_{N}}+\mathcal{O}(G_N^0).
\label{eq:area_mod_Ham}
\end{equation}
Here, Area denotes the area operator associated to the Ryu-Takayanagi surface anchored in the CFT subregion. By definition, the area is a local operator, and thus the modular Hamiltonian of a holographic CFT can always be approximated by the local expression \eqref{eq:area_mod_Ham} irrespective of whether a mapping from Rindler space exists or not. 

\paragraph{Modular Berry phase}
We now move on to discuss modular Berry phases as defined in \cite{Czech:2017zfq, Czech:2018kvg, Czech:2019vih}. In order to derive the modular Berry connection, we have to associate an algebra to the modular Hamiltonian. To this end, note that the area operator in  \eqref{eq:area_mod_Ham} and thus to leading order in $G_N$, the CFT modular Hamiltonian is the charge associated to
boosts near the edges of the RT surface in the dual bulk geometry \cite{Faulkner:2017vdd, Faulkner:2018faa}. From this point on, we therefore identify the modular Hamiltonian of the holographic CFT with the boost vector $K \rightarrow \xi^\mu\partial_\mu=H_{\mathrm{mod}} $ and refer to the boost vector as $H_{\mathrm{mod}}$. In this manner, the algebra we associate to the modular Hamiltonian is that generated by the boost generators. From the CFT point of view, these are the $SL(2,\mathbb{R})$ generators $L_{-1},L_{0}, L_1$. This was first employed in \cite{Czech:2017zfq,Czech:2019vih} to derive the modular Berry curvature for the vacuum CFT (see app. \ref{app:vacuum} for a summary).

\begin{figure}[t]
\vspace{-0.5cm}
\begin{center}
\includegraphics[width=0.45\linewidth]{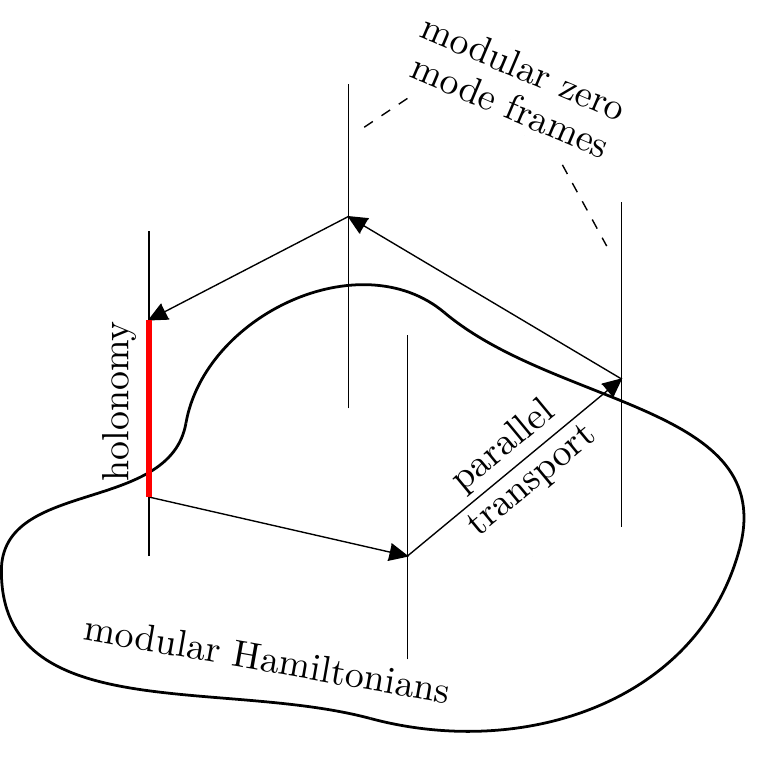}
\end{center}
\vspace{-0.8cm}
\caption{Illustration of modular transport. The base space consists of the modular Hamiltonians satisfying \eqref{eq:ModularHamiltonian}; the fibers are modular zero-mode frames. Due to the zero-mode ambiguity of $\partial_\lambda U^\dagger U$ in \eqref{eq:deformation_H}, parallel transport around a closed loop leads to a holonomy, known as the modular Berry phase.}
\label{fig:ModularTransport}
\vspace{-0.3cm}
\end{figure}

Modular Berry phases as defined in \cite{Czech:2017zfq, Czech:2018kvg, Czech:2019vih} are generated by operators $Q_a$ that commute with the modular Hamiltonian,
\begin{equation}
\left[H_{\mathrm{mod}},Q_a\right]=0.
\label{eq:ModularHamiltonian}
\end{equation}	
These operators $Q_a$, which we will refer to as modular  zero modes from now on, imply that there is a gauge ambiguity in defining operators in the subregion $A$: While expectation values of operators in subregion $A$ are invariant, the algebra is mapped to itself. The same applies to the complement $\bar{A}$, which has an independent gauge redundancy. Therefore, observers in $A$ and $\bar{A}$ have independent choices of reference frames by choosing a zero-mode frame, whereas the global state is sensitive to these relative choices of reference frame. After choosing a particular interval $\lambda^i=[u,v]$ to define the subregion $A$ on a constant time slice in the CFT, we deform $A$ infinitesimally, $\lambda^i +\delta \lambda^i$. Such a deformation implies that the modular Hamiltonian also changes, $H_{\mathrm{mod}}(\lambda^i)\rightarrow H_{\mathrm{mod}}(\lambda^i +\delta \lambda^i)$, since the deformation of the subregion implies that the reduced density matrix changes. To relate the zero-mode frames of different subregions, we must parallel transport $H_{\mathrm{mod}}(\lambda^i)$. If we parallel transport a subregion around an infinitesimal closed loop by choosing successive deformations that return the subregion to its original shape, the zero-mode ambiguity then gives rise to a Berry phase since in general $H_{\mathrm{mod}}(\lambda^i)$ in the initial region $A$  and in the region $A$ reached after a closed loop will differ by a zero-mode transformation. This construction is visualized in \figref{fig:ModularTransport}.

These Berry phases may be calculated as follows. 
Following \cite{Czech:2019vih}, it is convenient to decompose $H_{\mathrm{mod}}$ into its spectrum $\Delta$ and a basis $U$, $H_{\mathrm{mod}}=U^\dagger \Delta U$. Since in general, both $U$ and $\Delta$ will change under interval deformations $\delta \lambda$, the change of the modular Hamiltonian under the deformation is given by
\begin{equation}
\partial_\lambda H_{\mathrm{mod}}=\left[\partial_\lambda U^{\dagger} U, H_{\mathrm{mod}}\right]+U^{\dagger} \partial_\lambda\Delta U,
\label{eq:deformation_H}
\end{equation}
where the first term gives the change in the basis $U$ and the second one encodes the change in spectrum. The latter is 
an element of the zero-mode algebra, i.e. $[U^{\dagger} \partial_\lambda\Delta U, H_{\mathrm{mod}}]=0$. 

The Berry connection is given by the projection of the operator $\partial_\lambda U^{\dagger} U$ implementing the change in basis onto the zero-mode sector \cite{Czech:2019vih},
\begin{equation}
\Gamma\left(\lambda, \delta \lambda\right)=P_{0}^{\lambda}\left[\partial_{\lambda} U^{\dagger} U\right] \delta \lambda.
\label{eq:Berry_connection}
\end{equation} 
This may be seen as follows. Since we have not fixed a gauge in \eqref{eq:deformation_H}, the choice of basis $U$ is not unique; we also could have chosen $\tilde{U}=UU_Q$, where $U_Q$ is a unitary generated by a zero mode $Q_a$. Fixing the gauge by requiring $P_{0}^{\lambda}(\partial_\lambda \tilde{U}^{\dagger} \tilde{U})= 0$ then yields $\left(\partial_\lambda -P_{0}^{\lambda}\left[\partial_{\lambda} U^{\dagger} U\right]\right)U_Q=0$. Therefore, the Berry connection \eqref{eq:Berry_connection} follows. We now define a modular parallel transport operator $V_{\delta \lambda}= \tilde{U}^{\dagger}\partial_{\lambda}\tilde{U}$. This operator satisfies
\begin{equation}
\begin{aligned}
\partial_\lambda H_{\mathrm{mod}}&=\left[V_{\delta \lambda}, H_{\mathrm{mod}}\right]+P^{\lambda}_{0}(	\partial_\lambda H_{\mathrm{mod}})\\
P_{0}^{\lambda}(V_{\delta \lambda})&= 0.
\label{eq: equations modular Berry transport}
\end{aligned}
\end{equation}
The set of equations \eqref{eq: equations modular Berry transport} define modular Berry transport, visualized in \figref{fig:ModularTransport}.  By definition the modular parallel transport operator $V_{\delta \lambda}$ cannot have any zero-mode contributions. Hence, its zero-mode projection $P_{0}^\lambda$ is defined to vanish. Then, in order to calculate the modular Berry curvature, we first have to solve \eqref{eq: equations modular Berry transport} to find the generator of modular parallel transport and ensure that it does not have a zero-mode contribution. The modular Berry curvature operator follows from 
the commutator of two modular Berry transport generators,
\begin{equation}
\hat{R}_{ij}=[V_{\delta \lambda_i},V_{\delta \lambda_j}].
\label{eq:modular_Berry_curvature}
\end{equation}
It was shown in \cite{Czech:2019vih} that for an interval in the vacuum, the Berry curvature \eqref{eq:modular_Berry_curvature} is given in terms of the Crofton form on kinematic space (see app. \ref{app:vacuum} for a review).

\paragraph{Kinematic space}
Kinematic space in the context of holography was introduced in \cite{PhysRevD.90.106005}. On a constant time slice in the CFT, the kinematic space is the space of all CFT intervals. From the bulk point of view, kinematic space is then the space of all oriented RT surfaces that lie inside the bulk on the particular constant time slice. It is well known that RT surfaces calculate the entanglement entropy associated to the interval $[u,v]$ in the dual bulk geometry \cite{Ryu:2006bv},
\begin{equation}
S(u,v)=\frac{l(u,v)}{4G_N},
\end{equation}
where $l(u,v)$ is the length of the regularized shortest geodesic connecting  the interval endpoints $u,v$ in the bulk.

\begin{figure}[t]
\vspace{-0.3cm}
\begin{center}
\includegraphics[width=0.4\linewidth]{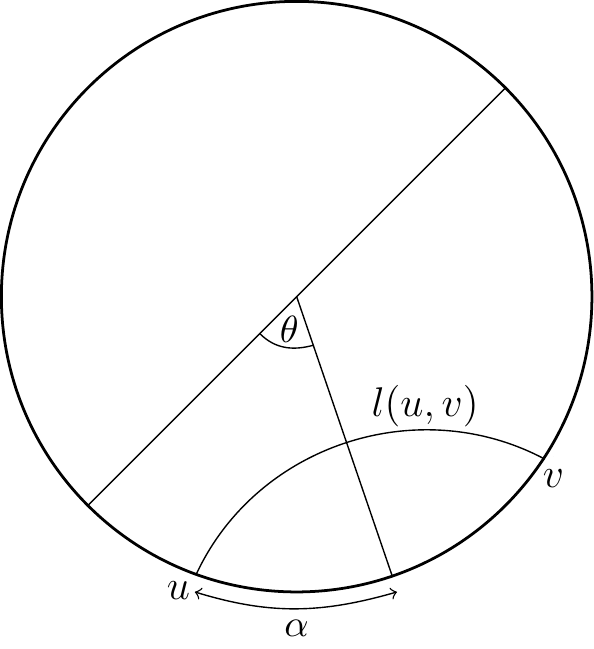}
\end{center}
\vspace{-0.8cm}
\caption{Coordinates of kinematic space. The disk and its enclosing circle represent a constant time slice of AdS spacetime and the boundary CFT, respectively. The RT surface with length $l(u,v)$ yields the entanglement entropy. An equivalent representation is given by exchanging $u$ and $v$ for the angular coordinates $\alpha$ and $\theta$.}
\label{fig:KinematicSpace}
\end{figure}
 Points in kinematic space are characterized by two variables, either the interval endpoints $[u,v]$ or the midpoint $\theta$ of the interval and the opening angle $\alpha$ (see \figref{fig:KinematicSpace}) of the RT surface anchored at $[u,v]$,
\begin{equation}
u=\theta-\alpha,\qquad v=\theta+\alpha.
\end{equation}

We will mostly employ $u,v$ coordinates. On kinematic space, we may then define a metric and a corresponding volume form induced by the entanglement entropy \footnote{Note that we define kinematic space in terms of the entanglement entropy. Therefore, kinematic space is defined by the shortest geodesics. In the literature, there also exist definitions in terms of all geodesics. Our notion of kinematic space and our conventions are inspired by \cite{Asplund:2016koz} on whose results for kinematic space in thermal CFTs we later rely.} \cite{Czech:2015qta},
\begin{align}
ds^2&=\frac{12}{c}\partial_{\lambda_i}\partial_{\lambda_j} S({\lambda_i},{\lambda_j})d{\lambda_i} d{\lambda_j},\\
\label{eq:relation_Crofton_entanglement}
\omega&=\frac{12}{c}\partial_{\lambda_i}\partial_{\lambda_j} S({\lambda_i},{\lambda_j})d{\lambda_i} \wedge d{\lambda_j}.
\end{align}
The volume form on kinematic space is referred to as the Crofton form. Furthermore, locally $\omega$ can be expressed in terms of a symplectic potential that is given in terms of the differential entropy \cite{Czech:2015qta},
\begin{equation}
\omega = d\kappa \qquad \text{with}\qquad \kappa=-\partial_{\lambda^i}S|_{\lambda^{j}}.
\end{equation}

We demonstrate on various examples in the next section that the relation between the Crofton form on kinematic space and the Berry curvature observed for the vacuum in \cite{Czech:2019vih} also holds for intervals on a constant time slice in a thermal CFT. We show that this has particularly interesting consequences when there are transitions in the Crofton form. These transitions in the Crofton form occur in certain thermal CFTs if there is a transition in the entanglement entropy, see e.g. \cite{Asplund:2016koz}. 

\subsection{Modular Berry curvature for a thermal CFT}
\label{sec:Berry curvature thermal}
Here, we present new results for modular Berry phases for intervals in CFTs in a thermal state. To obtain the modular Berry curvature, we proceed as described in \cite{Czech:2019vih} for the vacuum. For completeness, we have included a review in app. \ref{app:vacuum}. The best approach to solve \eqref{eq: equations modular Berry transport} is to find eigenoperators $E_k$ of the adjoint action of the modular Hamiltonian \cite{Czech:2019vih},
\begin{equation}
[E_{k}, H_{\mathrm{mod}}]=\kappa E_{k}.
\end{equation} 
Possible solutions to this equation in the cases we consider throughout this paper are
\begin{equation}
\begin{aligned}
E_{k,1}&= H_{\mathrm{mod}},\\
E_{k,2}&=\partial_\lambda H_{\mathrm{mod}}.
\end{aligned}
\end{equation}
The first solution $E_{k,1}$ belongs to the zero-mode sector since it has eigenvalue $\kappa=0$. Such contributions cannot contribute to the generator of modular Berry transport to ensure \eqref{eq: equations modular Berry transport} holds true. The second solution $E_{k,2}$ has a non-vanishing eigenvalue $\kappa$ in the examples we consider. Therefore, the generator of modular Berry transport can be constructed from $E_{k,2}$,
\begin{equation}
V_{\delta \lambda}=\frac{1}{\kappa }E_{k,2}=\frac{1}{\kappa }\partial_\lambda H_{\mathrm{mod}}.
\end{equation}
Then, we insert this solution into \eqref{eq:modular_Berry_curvature} to obtain the modular Berry curvature operator. Note that this commutator can be simplified using standard identities such that the Berry curvature is simply given by (see also \cite{Huang:2020cye} for a derivation)
\begin{equation}
\hat{R}=-\frac{2}{\kappa}\partial_{\lambda^i}\partial_{\lambda^j}H_{\mathrm{mod}}\delta \lambda^{i}\wedge \delta\lambda^{j}.
\label{eq:Berry_curvature_Simple}
\end{equation}

In the thermal configurations we consider, the result for the modular Berry curvature operator is given by
\begin{equation}
\hat{R}=\omega H_{\mathrm{mod}} .
\label{eq:relation_Berry curvature_Crofton form}
\end{equation}
As the modular Berry curvature is proportional to $H_{\mathrm{mod}} $, it is therefore a zero-mode itself and scales with a particular prefactor $\omega$. We show that $\omega$ is the Crofton form on kinematic space whenever intervals are chosen to lie on a constant time slice for the configurations in thermal CFTs we consider. Therefore, our results for thermal CFTs and those of \cite{Czech:2019vih} for the vacuum support that the modular Berry curvature and the entanglement entropy for intervals that lie on a constant time slice are related as follows,
\begin{equation}
\langle \hat{R}\rangle =\langle -\frac{2}{\kappa}\partial_{\lambda^i}\partial_{\lambda^j}H_{\mathrm{mod}}\rangle d\lambda^i \wedge d\lambda^j=\omega  \langle H_{\mathrm{mod}} \rangle=\frac{12}{c}\partial_{\lambda^i}\partial_{\lambda^j} S(\lambda^i,\lambda^j)\langle H_{\mathrm{mod}} \rangle d\lambda^i \wedge d\lambda^j.
\label{eq:relation_entanglement}
\end{equation}

The relation \eqref{eq:relation_entanglement} is particularly interesting in thermal CFTs since in those systems the entanglement entropy is known to exhibit transitions in the configuration yielding the shortest RT geodesic. Then, \eqref{eq:relation_entanglement} indicates that such transitions also occur in the modular Berry curvature. We discuss these transitions for a single interval in a BTZ black hole spacetime and two intervals, one on each boundary, for the eternal AdS black hole. For the latter, we observe that when both intervals are put on shifted time slices before the transition occurs, a spacetime wormhole occurs with a Berry phase related to the time shift.

\subsubsection{Modular Berry curvature from thermal density matrix}
\label{subsec:thermal}
We begin with the simplest configuration for a CFT in a thermal state. When the subregion on the CFT is chosen to encompass the full constant time slice, the density matrix is given in terms of the thermal density matrix and the modular Hamiltonian is simply given by the system Hamiltonian, $H_{\mathrm{mod}}=K_++K_-=2\beta H$. Following app. \ref{app:vacuum}, the modular Hamiltonian may then be written as\footnote{We only consider the holomorphic sector, the antiholomorphic sector is analogous.}
\begin{equation}
K_+=2\beta L_0.
\end{equation} 
In this case, we find that the Berry curvature vanishes,
\begin{equation}
R=0.
\label{eq:Berry_curvature_thermal_state}
\end{equation}
In such a system, it is not possible to construct a modular Berry transport operator $\partial_\lambda \tilde{U}^\dagger \tilde{U}$ that solves \eqref{eq: equations modular Berry transport}. In particular, \eqref{eq: equations modular Berry transport} implies that for $K_{+}=2\beta L_0$, we have to solve $0=[\partial_\lambda \tilde{U}^\dagger \tilde{U},2\beta L_0]$. The only possible solution is $\partial_\lambda \tilde{U}^\dagger \tilde{U}=f(u,v) L_0$, where we can choose an arbitrary, not necessarily constant function $f$. However, this operator is an eigenoperator to the adjoint action of $K_+$ with eigenvalue zero and is thus a zero mode. Therefore, $P^\lambda_{0} (\partial_\lambda \tilde{U}^\dagger \tilde{U})=f(u,v) L_0\neq0$, and the second equation in \eqref{eq: equations modular Berry transport} is no longer true. 

Since for a CFT in a thermal state the modular Hamiltonian is given in terms of the system Hamiltonian, an observer has access to the full system. In contrast, when we consider the modular Hamiltonian for a finite interval, the division of the system into two or more subregions implies that the observer only has access to the degrees of freedom in the subregion associated to the CFT interval. The entanglement cut effectively acts as a horizon hiding the remaining degrees of freedom, including zero modes, in the complement to the interval subregion. This introduces independent gauge redundancies for observers in the subregions and its complement. Both observers may independently choose the basis $U$ for their modular Hamiltonian and thereby fix their zero-mode frame. The global state is sensitive to this relative choice of zero-mode frame and picks up additional phases. When the interval is now chosen to be so large that it effectively encompasses the full spatial region, there no longer is an independent choice of gauge. The observer has access to the full system and fixes the gauge globally. We then can no longer obtain a Berry phase.

\subsubsection{Modular Berry curvature for a finite  interval in the cylinder} 
\label{subsec:black_string}
The modular Berry curvature can be calculated in a similar manner for a finite interval $[u,v]$ in a thermal CFT on the cylinder with compactified Euclidean time direction with radius $\frac{\beta}{2}$. This corresponds to a BTZ black string in the bulk, i.e. a black hole with non-compact spatial direction. 

The modular Hamiltonian for an interval $[u,v]$ in a thermal CFT on the cylinder is given in \cite{Hartman:2015apr} and may be written as
\begin{equation}
\begin{aligned}
K_+&=s_1L_1+s_0L_0+s_{-1}L_{-1},\\
s_1=\frac{2 \beta \coth \frac{2\pi}{\beta}\left(v-u\right) / 2}{e^{ \frac{2\pi}{\beta}u}+e^{ \frac{2\pi}{\beta}v}},\quad
s_0&=-2\beta  \coth \frac{2\pi}{\beta}\frac{v-u}{2},\quad
s_{-1}=\frac{2 \beta \coth \frac{2\pi}{\beta}\left(v-u\right) / 2}{e^{- \frac{2\pi}{\beta}u}+e^{ -\frac{2\pi}{\beta}v}}.
\end{aligned}
\label{eq:new_thermal_mod_Hamiltonian}
\end{equation}
We now solve \eqref{eq: equations modular Berry transport} and obtain
\begin{align}
V_{\delta u}&=\frac{1}{2\beta }\partial_u K_+,\\
V_{\delta v}&=-\frac{1}{2\beta }\partial_v K_+.
\end{align}
Note that $P^{\lambda}_{0}(	\partial_\lambda H_{\mathrm{mod}})=0$ as for the vacuum because the operator $\partial_\lambda \tilde{U}^\dagger U$  is an eigenoperator of the adjoint action of the modular Hamiltonian with non-vanishing eigenvalue $\pm \frac{1}{2\beta}$.
The Berry curvature operator then reads
\begin{equation}
\hat{R}=\frac{1}{2\beta } \partial_u\partial_v K_+ du\wedge dv=-\frac{1}{2\beta}\frac{4\pi^2}{\beta^2} \mathrm{csch}^2 \frac{2\pi }{\beta }\frac{v-u}{2} K_+du\wedge dv.
\label{eq:Berry_curvature_thermal_1}
\end{equation}
The Berry curvature is non-vanishing in contrast to \eqref{eq:Berry_curvature_thermal_state}. The interval divides the system into two subregions such that both regions become entangled. Here, the entanglement cut acts like a horizon, hiding some of the degrees of freedom, including zero modes behind the entanglement cut. The full gauge symmetry of the system is broken into a much smaller subgroup. The Berry phase arises by an independent choice of modular zero-mode frame that each observer in the subregion and its complement may choose, which gives rise to a relative zero-mode phase in the global state of the system. Furthermore, the results \eqref{eq:Berry_curvature_thermal_1} is consistent with \eqref{eq:Berry_curvature_thermal_state}. The limit, for which the modular Hamiltonian is given by the system Hamiltonian, $H_{\text{mod}}=2\beta H$,  corresponds to taking the large interval limit $u,v,(v-u)\gg\beta$  \cite{Hartman:2015apr}. In this case, there is no entanglement between subsystems. In this limit, the Berry curvature \eqref{eq:Berry_curvature_thermal_1} vanishes as expected. 

Moreover, the Crofton form on kinematic space for the BTZ black string was derived in \cite{Asplund:2016koz},
\begin{equation}
	\omega=\frac{4 \pi^{2}}{\beta^{2}} \frac{1}{\sinh ^{2}\left(\frac{\pi(v-u)}{\beta}\right)} du\wedge dv.
\end{equation}
By comparison with \eqref{eq:Berry_curvature_thermal_1}, we find that the Berry curvature is given in terms of the Crofton form on the kinematic space of the BTZ black string. This result indicates that the relation between the Crofton form and the Berry curvature also holds for thermal CFTs and supports the relation \eqref{eq:relation_entanglement} for intervals in a holographic CFT on a constant time slice. Furthermore, since the entanglement entropy does not exhibit a transition for the BTZ black string, we find only one Berry curvature and thus one Berry phase independent of the interval size. Note, however, that the Berry phase vanishes when the interval encompasses the full constant time slice in accordance with the results of \secref{subsec:thermal}.
\subsubsection{Two disjoint intervals in a thermal CFT on the cylinder}
\label{subsec:2-intervals}
We now consider an eternal AdS black hole with two CFTs at the asymptotic boundaries and choose an interval in each CFT, $[u_L,v_L]$ and $[u_R,v_R]$. This corresponds to a setup with two disjoint intervals in thermal CFTs. Such systems are particularly interesting because a transition in the entanglement entropy depending on the interval size was derived in \cite{Hartman:2013qma}. Furthermore, for the eternal AdS black hole, there is a wormhole in the bulk. For particular choices of intervals in the boundary CFT, we then expect to obtain non-local terms in the Berry curvature that otherwise do not arise when studying single-sided systems.
 
\begin{figure}[t]
\vspace{-0.5cm}
\begin{center}
\includegraphics[width=0.6\linewidth]{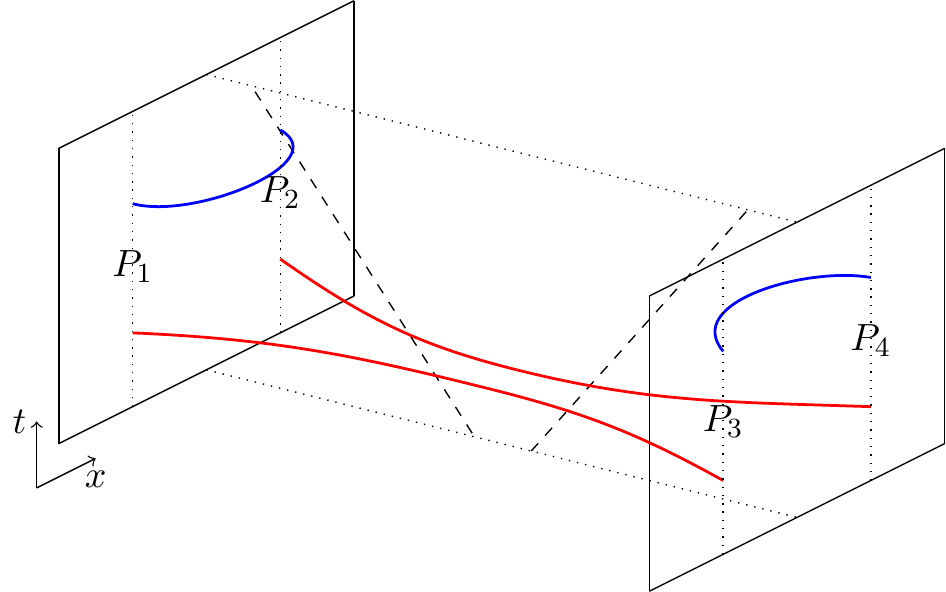}
\end{center}
\vspace{-0.8cm}
\caption{Two configurations of RT geodesics connecting two intervals in the CFT in the presence of an eternal black hole. The two solid rectangles are regions of the left and right boundary CFTs. In the bulk, the dashed lines represent the horizon of the eternal black hole. The dotted lines are included to guide the eye of the reader. The coloured lines correspond to the geodesics associated to the endpoints $P_i$, drawn for late and early times. For $\kappa>1$, $\kappa$ given by \eqref{eq:kappa}, the shortest geodesics connect points in opposite boundaries, thereby stretching through the wormhole (red lines). At late times when $\kappa\leq 1$, the shortest geodesics connect endpoints in the same boundary and do not cross the horizon (blue line).}
\label{fig:RT_Transition}
\end{figure}

Based on \cite{Nakagawa:2018kvo}, we employ the following modular Hamiltonians for two disjoint intervals in the CFT,
\begin{equation}
\begin{aligned}
&\mathrm{for}\, \kappa\leq1:\\
&K_{+,L}=\frac{2 \beta \coth \frac{2\pi}{\beta}\left(v_L-u_L\right) / 2}{e^{ \frac{2\pi}{\beta}u_L}+e^{ \frac{2\pi}{\beta}v}}L_1-2\beta  \coth \frac{2\pi}{\beta}\frac{v_L-u_L}{2}L_0+\frac{2 \beta \coth \frac{2\pi}{\beta}\left(v_L-u_L\right) / 2}{e^{- \frac{2\pi}{\beta}u_L}+e^{ -\frac{2\pi}{\beta}v_L}}L_{-1} \\ 
&\mathrm{for}\, \kappa>1:\\
& K_{+,(u_L,u_R)}=\frac{2 \beta \coth \frac{2\pi}{\beta}\left(u_L-u_R\right) / 2}{e^{ \frac{2\pi}{\beta}u_L}+e^{ \frac{2\pi}{\beta}u_R}}L_1-2\beta  \coth \frac{2\pi}{\beta}\frac{u_L-u_R}{2}L_0+\frac{2 \beta \coth \frac{2\pi}{\beta}\left(u_L-u_R\right) / 2}{e^{- \frac{2\pi}{\beta}u_L}+e^{ -\frac{2\pi}{\beta}u_R}}L_{-1} ,
\end{aligned}
\end{equation}
where 
\begin{equation}
\kappa=\frac{\sinh \frac{2\pi}{\beta}\frac{u_L-v_R}{2}\sinh \frac{2\pi}{\beta}\frac{u_R-v_R}{2}}{\textrm{csch} \frac{2\pi}{\beta}\frac{u_L-u_R}{2}\textrm{csch} \frac{2\pi}{\beta}\frac{v_L-v_R}{2}}.\label{eq:kappa}
\end{equation}
$K_{+,R}$ and $K_{+,(v_L,v_R)}$ are obtained by replacing the interval endpoints with their appropriate counterparts.
The transition between the modular Hamiltonians occurs when $\kappa=1$ and is caused by a transition in RT geodesic configurations in the bulk at the same value $\kappa=1$ \cite{Hartman:2013qma}. For $\kappa > 1$, the shortest geodesic is one that connects interval endpoints at both boundaries. Therefore, the RT geodesic stretches through the wormhole in the bulk. For a visualization see \figref{fig:RT_Transition}. The modular Hamiltonian is then $K_{+,(u_L,u_R)}$ and similarly for $v$. For such a configuration, we again expect non-local contributions due to the presence of the wormhole. In contrast, for $\kappa\leq1$, the RT geodesic sits outside the horizon and connects interval endpoints on the same boundary. The modular Hamiltonian is  $K_{+,L}$ and similarly for the other boundary. In this case, we do not expect the modular Berry curvature to feel the presence of the wormhole and expect a local result related to kinematic space.

Proceeding as described in the previous sections, for deformations $u_L+\delta u_L$ and $u_R+\delta u_R$ and similarly for $v_L$, the Berry curvature is given by
\begin{equation}
\begin{aligned}
\hat{R}&=\frac{1}{2\beta} \partial_u\partial_v K_{+,L} du_L\wedge dv_L+\frac{1}{2\beta} \partial_u\partial_v K_{+,R} du_R\wedge dv_R\\&=-\frac{1}{2\beta}\frac{4\pi^2}{\beta^2}\left( \operatorname{csch}^{2} \frac{2 \pi(\frac{v_L-u_L}{2}) }{\beta}K_{+,L} du_L \wedge dv_L\right.+\left. \operatorname{csch}^{2} \frac{2 \pi(\frac{v_R-u_R}{2}) }{\beta}K_{+,R} du_R \wedge dv_R\right)
\end{aligned}
\end{equation}
for $\kappa\leq 1$
and 
\begin{equation}
\begin{aligned}
\hat{R}&=-\frac{1}{2\beta}\frac{4\pi^2}{\beta^2}\left( \operatorname{csch}^{2} \frac{2 \pi(\frac{u_R-u_L}{2}) }{\beta}K_{+,(u_L,u_R)}du_L \wedge du_R\right.\left.+ \operatorname{csch}^{2} \frac{2 \pi(\frac{v_R-v_L}{2}) }{\beta}K_{+,(v_L,v_R)} dv_L \wedge dv_R\right)
\end{aligned}
\end{equation}
for $\kappa >1$.
For an independent interval choice on the left and right boundary, we thus obtain a non-vanishing modular Berry curvature that exhibits a transition. To make explicit the non-local contributions to the modular Berry curvature for RT geodesics that connect both boundaries, it is convenient to choose specific interval endpoints. In line with \cite{Hartman:2013qma}, we choose the points $P_1=(-\frac{x}{2},t_L=-t)$, $P_2=(\frac{x}{2},t_L=-t)$ in the left boundary and $P_3=(-\frac{x}{2},t_R=t)$ and $P_4=(\frac{x}{2},t_R=t)$ in the right boundary, where we have identified the Schwarzschild time in the left and right boundary by $t_L=t_R+\frac{i\beta }{2}$.
Note that by establishing a relation between the time in the left and right boundary, we have identified the modular parameters on both sides. For this choice of interval endpoints, the transition at $\kappa=1$ occurs at $t=\frac{x}{2}$. The entanglement entropies are given by \cite{Hartman:2013qma}
\begin{equation}
\begin{array}{l}
S=\frac{2 c}{3} \log \left(\cosh \frac{2 \pi t}{\beta}\right)+S_{\mathrm{div}} \\
S=\frac{2 c}{3} \log \left(\sinh \frac{\pi x}{\beta}\right)+S_{\mathrm{div}}.
\end{array}
\end{equation}
 Thus, for $t<\frac{x}{2}$, we connect the endpoints $P_1$ and $P_3$ in each boundary CFT with a RT geodesic through the wormhole. The interval is then given by $[u_L,v_L]=[-t+i\frac{\beta }{2},t]$ at fixed position $-\frac{x}{2}$ and similarly for $[u_R,v_R]$. After the transition, the interval endpoints are given by $P_1$ and $P_2$, which amounts to an interval $[u_L,v_L]=[-\frac{x}{2},\frac{x}{2}]$ at fixed time $-t+\frac{i\beta}{2}$ and similarly for the remaining points. For the latter setup, we simply obtain twice the Berry curvature we have obtained for the one-interval setup in \eqref{eq:Berry_curvature_thermal_1},
\begin{equation}
R=-2\frac{1}{2\beta}\frac{4\pi^2}{\beta}\mathrm{csch}^2\frac{2\pi x}{\beta}dx_1\wedge dx_2,
\end{equation}
where we have labeled the interval endpoints, which are identical on both sides, $x_1$ and $x_2$ to demonstrate that these points are to be considered independent. This is the expected result: As the RT geodesic stays in each exterior region, it is not sensitive to the wormhole in the bulk and we obtain twice the modular Berry curvature of the single-sided setup. In the large interval limit, the Berry curvature vanishes. 

Since we identify the time in the left and right exterior region, whenever we consider the same timeslices in the left and right CFT before the transition, we obtain zero Berry curvature due to $dt\wedge dt=0$. This is consistent with a wormhole interpretation of this Berry phase: If the time in the left and right boundary are identified, there is no relative zero-mode frame in the global TFD state and the Berry phase vanishes. We may, however, obtain a Berry phase by introducing an additional shift $\delta$, i.e. rather than $t_L=t_R+\frac{i\beta}{2}$, we set $t_L-\delta=t_R+\frac{i\beta}{2}$. The time in the left and right CFT are then identified only up to a shift. This is also consistent with a wormhole interpretation of the bulk spacetime. The presence of the wormhole prohibits the global identification of the left and right boundary since there is no global timelike Killing vector \cite{Verlinde:2020upt}. To relate the times in both exterior regions, a shift $\delta$ is then introduced. This shift is non-local since it cannot be measured by a single observer \cite{Verlinde:2020upt}. Before the transition, the interval is now given by $[-t-\delta+\frac{i\beta}{2},t]$. We then obtain the modular Berry curvature
\begin{equation}
\hat{R}_{u_L,u_R}=-\frac{1}{2\beta}\frac{4\pi^2}{\beta^2}\mathrm{sech}^2\left(\frac{\pi}{\beta}(2t+\delta)\right)K_{+,(u_L,u_R)}d\delta \wedge dt.
\end{equation}
On the other hand, shifting both interval endpoints $t_L\rightarrow t_L-\delta =-t-\delta $ and $t_R\rightarrow t_R-\delta=t-\delta$ yields
\begin{equation}
\begin{aligned}
\hat{R}_{u_L,u_R}&=\frac{1}{2\beta}\frac{4\pi^2}{\beta^2}\mathrm{sech}^2\left(\frac{2\pi}{\beta}t\right)K_{+,(u_L,u_R)}(-dt-d\delta )\wedge (dt-d\delta)\\
&=-\frac{1}{2\beta}\frac{4\pi^2}{\beta^2}\mathrm{sech}^2\left(\frac{2\pi}{\beta}t\right) 2K_{+,(u_L,u_R )}d\delta \wedge dt.
\end{aligned}
\end{equation}
Note that for a shift in opposite directions $t_L\rightarrow t_L-\delta =-t-\delta $ and $t_R\rightarrow t_R+\delta=t+\delta$, we obtain
\begin{equation}
R=0.
\end{equation}
In this case, the Berry curvature vanishes because shifting in this manner corresponds to a time evolution with $H_L-H_R$, which is a symmetry of the system. Such an evolution cannot lead to a Berry phase. 

We interpret these results as follows. If the modular Berry curvature is given in terms of local variables ($\kappa \leq 1$), the Ryu-Takayanagi surface does not stretch far into the bulk. Therefore, the entanglement wedges associated to the RT surfaces anchored on the left and right boundary do not reach deep into the bulk. For such a configuration, IR physics is irrelevant. On the other hand, for $\kappa >1$, the RT surfaces connect both boundaries and consequently the entanglement wedges overlap. 
However, due to the presence of a horizon in the bulk, there is no global time coordinate and the time in the left and right exterior region are identified only up to a shift $\delta$. This gives rise to a spacetime wormhole Berry curvature due to the variable $\delta$. 

We stress that though the modular Berry phase exhibits the same features we expect for Berry phases related to spacetime wormholes discussed in sec. \ref{sec:Virasoro_Berry_phase} and \ref{sec:gauge_Berry_phase}, the Berry phase is nevertheless different in origin from that in those sections. Here, we consider modular Berry phases, which arise from deformations of intervals in the CFT. These are sometimes referred to as shape-changing Berry phases \cite{deBoer:2021zlm}. On the other hand, in sec. \ref{sec:Virasoro_Berry_phase}, we consider the full CFT without subregions and apply transformations that change the state. The relative shift $\delta$ in the boundary time gives rise to the modular Berry phase if the RT surface stretches through the horizon. This shift also occurs in sec. \ref{sec:gauge_Berry_phase}, where we consider Berry phases from gauge transformations. Nevertheless, the modular Berry phases are very different from gauge Berry phases as for the first we consider shape deformations, whereas for the latter gauge transformation in the asymptotic boundary region are relevant.      

\subsubsection{Modular Berry curvature for a finite interval on the torus}
\label{subsec:thermal_CFT_1_sided}
	Here, we argue that for holographic CFTs on the torus dual to a BTZ black hole, the Berry curvature exhibits a transition induced by a similar transition in the entanglement entropy. We further comment that the knowledge of the Crofton form and its relation to the Berry curvature in holographic CFTs provides a means to calculate modular Hamiltonians for holographic CFTs.
	
	Upon compactifying the spatial coordinate on the cylinder, we obtain a thermal CFT on a torus, which is dual to a BTZ black hole in the bulk. For convenience, we set the spatial radius of the torus to $r=1$. The temperature is given in terms of the horizon $r_H$, $\beta=\frac{2\pi}{r_H}$. In such systems, the entanglement entropy exhibits a transition. For small intervals, $\frac{v-u}{2}<\alpha_c$, where $\alpha_c$ denotes the critical opening angle,
	\begin{equation}
	\alpha_c=\frac{\beta}{4 \pi } \log \left(\frac{1}{2}+\frac{1}{2} \exp \left(\frac{4 \pi^{2} }{\beta}\right)\right),
	\end{equation}  
	 the entanglement entropy in the bulk is as usual given by the shortest geodesic connecting the interval endpoints. 
	After the transition for intervals $\frac{v-u}{2}>\alpha_c$, the shortest geodesic is no longer given by a single geodesic connecting both interval endpoints but rather by two disconnecting geodesics, one connecting the interval endpoints with the shortest geodesics for the complement but with reverse orientation and a second horizon-wrapping geodesic. For a detailed discussion, we refer to \cite{Asplund:2016koz}. The Crofton form on kinematic space is given by   
		\begin{equation}
	\begin{aligned}
	\omega_1&=\frac{4 \pi^{2}}{\beta^{2}} \operatorname{csch}^{2} \frac{\pi(v-u)}{\beta} du\wedge dv,\\
	\omega_2 &=-\frac{4 \pi^{2}}{\beta^{2}} \operatorname{csch}^{2} \frac{2 \pi(\pi-\frac{v-u}{2}) }{\beta} du \wedge dv,
	\end{aligned}
	\label{eq:transition_Crofton}
	\end{equation}
	where $\omega_1$ and $\omega_2$ are the Crofton forms before and after the transition, respectively. At the transition, the Crofton form is singular \cite{Asplund:2016koz}. 
	
	In general, it is difficult to derive the Berry curvature for CFTs on the torus since the form of the modular Hamiltonian becomes theory-dependent and the modular Hamiltonian cannot be written in the form \eqref{eq:mod_Ham_local} \cite{Cardy:2016fqc}. Typically, known results for the modular Hamiltonian are restricted to free CFTs such as for free fermions on the torus in \cite{Blanco:2019cet, Fries:2019ozf} and in \cite{Fries:2019ozf} also for multiple intervals. 
	 However, for a holographic CFT, the relation \eqref{eq:area_mod_Ham} between the boundary modular Hamiltonian and the bulk has been proposed to leading order in $G_N$.
	This suggests that for holographic CFTs, the modular Hamiltonian $K$ must always have a local contribution given in terms of the area operator, which defines the RT surface calculating the entanglement entropy in the bulk. In particular, the modular Hamiltonian may be written in the form \eqref{eq:mod_Ham_local} \cite{Jafferis:2015del} such that we may employ the procedure introduced in the previous sections to calculate the Berry curvature. 
	Before the transition occurs, the following modular Hamiltonian for an interval in a holographic CFT on the torus has been suggested in \cite{Asplund:2016koz},
	\begin{equation}
	    	K_{\mathrm{local},1}=2 \beta \int_{u}^{v} d \theta^{\prime} \frac{\sinh \frac{\pi(\theta^{\prime}-u)}{\beta} \sinh \frac{\pi(v-\theta^{\prime})}{\beta}}{\sinh \frac{2 \pi }{\beta}(\frac{u-v}{2})} T_{00}\left(\theta^{\prime}\right).
	    	\label{eq:H_torus_1}
	\end{equation}
	No expression for the modular Hamiltonian after the transition in the entanglement entropy was given in \cite{Asplund:2016koz} since the effect of the transition on the modular Hamiltonian was not clear. Let us first calculate the Berry curvature for the modular Hamiltonian before the transition. We may again identify $K\rightarrow \xi^\mu\partial_\mu$. Following the same steps as before, we set $\xi^\mu\partial_\mu\equiv H_{\mathrm{mod}}$ and write $H_{\mathrm{mod}}$ in terms of $SL(2,\mathbb{R})$ generators. Then, the Berry curvature may be obtained from the procedure introduced in \secref{sec:Berry curvature thermal}. We obtain 
	\begin{equation}
	    \hat{R}=\omega_1 H_{\mathrm{mod}},
	\end{equation}
	where $\omega_1$ is the Crofton form \eqref{eq:transition_Crofton} before the transition. This result is in agreement with the expected relation \eqref{eq:relation_entanglement} and is evidence that the modular Hamiltonian for a holographic CFT on the torus before the transition in the entanglement entropy is indeed given by \eqref{eq:H_torus_1}. Furthermore, \eqref{eq:relation_entanglement} implies that the Berry curvature should exhibit the same transition as the Crofton form. Therefore, after the transition, the Berry curvature must be of the form
	\begin{equation}
	    \hat{R}=\omega_2 H_{\mathrm{mod}},
	    \label{eq:B_curvature_after_transition}
	\end{equation}
	where $\omega_2$ is the Crofton form after the transition given in \eqref{eq:transition_Crofton}. It is now possible to obtain the modular Hamiltonian after the transition that gives the correct Berry curvature \eqref{eq:B_curvature_after_transition},
	\begin{equation}
	    K_{\mathrm{local},2}=2 \beta \int_{u}^{v} d \theta^{\prime} \frac{\sinh \frac{\pi\left(\pi-(\theta^{\prime}-u)\right)}{\beta} \sinh \frac{\pi\left(\pi-(v-\theta^{\prime})\right)}{\beta}}{\sinh \frac{2 \pi }{\beta}(\pi-\frac{u-v}{2})} T_{00}\left(\theta^{\prime}\right).
	\end{equation}
	The shift by $\pi$ in the argument of $\mathrm{sinh}$ compared to \eqref{eq:H_torus_1} reflects that after the transition the shortest geodesic is given in terms of the interval complement. 
	Therefore, at least for holographic CFTs, the modular Hamiltonian for intervals on a constant time slice may also be obtained from the Crofton form and its relation to the Berry curvature.

\section{Summary and Outlook}\label{sec:summary_and_conclusion}

In this work, we showed that the presence of wormholes in bulk geometries is linked to Berry phases in the holographic CFTs. We extended the connection between Berry phase and entanglement from simple quantum mechanical systems discussed in \cite{Nogueira:2021ngh} to quantum field theories that have gravity dual descriptions. The connection between wormholes in bulk geometries and Berry phases in the dual CFTs arises from mathematical properties of the Hilbert space that are realized both in the gravity theory and in its dual CFT. This connection is expected to play a pivotal role in understanding how the holographic correspondence emerges as a consequence of spacetime entanglement. 

We established a classification of the Berry phases, and their holographic description in terms of spacetime wormholes, by their generating diffeomorphisms. Coupling two Virasoro coadjoint orbits via an annulus times time geometry, we obtain a Berry phase arising from an improper diffeomorphism generating bulk geometries with different defects or ``Virasoro hair''. The holonomy on the annulus is associated with a wormhole. Furthermore, proper diffeomorphisms, which correspond to an asymptotic gauge symmetry, give rise to a Berry phase and associated wormhole when two CFTs are entangled. Finally, we calculated modular Berry phases for subregions in thermal CFTs, showing that the Berry curvature is given in terms of the Crofton form on kinematic space. Using this, we find that for two subregions in entangled thermal CFTs the Berry curvature displays a transition similar to that in the entanglement entropy \cite{Hartman:2013qma}. Let us now conclude with two immediate and concrete applications of the formalism we developed in this work. 

\paragraph{Non-factorization of the Hilbert space}
In \secref{sec:entangled-CFT} and \secref{sec:gauge_Berry_phase} we discussed how the corresponding types of Berry phases are related to non-factorization of the Hilbert space. Starting with Virasoro Berry phases, we made use of the results of \cite{Henneaux:2019sjx}, where it was shown that the boundary action of AdS$_3$ on an annulus topology consists of two chiral boson actions coupled by the holonomy $k_0$. The holonomy results from non-contractible loops in the annulus topology. Due to this coupling, the boundary actions only factorize when the holonomy is set to zero. In light of \cite{Verlinde:2021kgt}, we found that this manifests in the symplectic form \eqref{eq:symplectic form_BTZ} of the boundary phase space: the holonomy enters as a non-local term, which implies that the symplectic form is non-exact.

Moreover, using that the chiral boson action can be rewritten into the difference of two geometric actions on coadjoint orbits of the Virasoro group, we discussed how the non-factorization can also be understood in terms of Virasoro Berry phases. The orbit labels of these geometric actions are equal and related to the holonomy according to \eqref{eq:orbit_label}. As explained in \secref{sec:generalizing}, the symmetry underlying these two geometric actions factorizes into products associated with each boundary only when the holonomy is set to zero.

Although gauge Berry phases arise from a different type of diffeomorphisms and have different underlying symmetries, they have a similar interpretation to Virasoro Berry phases in terms of factorization. As discussed in \secref{sec:gauge_Berry_phase}, gauge Berry phases arise in presence of a horizon from transformations in the asymptotic symmetry group. As an example, we discussed the eternal black hole in AdS spacetime, dual to the TFD state. In presence of the horizon, there is no globally defined time. The associated ambiguity in relating the times on the left and right boundaries leads to introducing the variable $\delta$ \cite{Verlinde:2020upt}, w.r.t.~which a Berry connection and corresponding Berry phase can be defined \cite{Nogueira:2021ngh}. If there were no horizon, there would not be the above mentioned ambiguity and a gauge Berry phase could not be defined. Therefore, a gauge Berry phase signals non-factorization of the left and right boundaries.

We point out here that the above described phenomena have an interesting relation with the factorization map. Such a map is used in quantum field theories or quantum gravity to factorize the full Hilbert space. This enables to define quantities like entanglement entropy between the two smaller Hilbert spaces. For JT-gravity, this map is extensively studied in \cite{Jafferis:2019wkd}. It is found that the factorization map in two-dimensional gravity cannot be stated in terms of a local boundary condition for the path integral, but is related to the Euler character via the Gauss-Bonnet theorem. While the study was performed for JT gravity, the general features of a non-local boundary condition are expected to hold for gravity in higher dimensions as well. In our above discussion, we saw that using the non-local content of the theory, i.e. the holonomy $k_0$, the action \eqref{eq:action:non_Abelian} can be written in a factorized structure in terms of the actions on the inner and outer boundaries \eqref{eq:action_split}. Similarly, the entanglement between the coupled spins in \cite{Nogueira:2021ngh} is calculated by tracing out one of the $\mathbf{CP}^1$ factors, that is by imposing some factorization map.

\paragraph{Towards subregion complexity} 

As for a very different application, the results obtained in \secref{sec:modular_Berry_phase} of our present work may open up an interesting avenue to understand subregion complexity, using the CFT set ups considered in this work. Modular Berry phases and in particular the modular Hamiltonian are known to be a promising new tool to study subregion complexity \cite{Alishahiha:2015rta}. However, this notion of complexity has so far been studied mostly from the bulk point of view and a precise CFT interpretation of the same remains elusive. In light of the results presented here and in \cite{Czech:2019vih}, the relation between modular Berry curvature and the Crofton form suggests that the complexity=volume proposal for spatial subregions $\Sigma$ contained beneath the RT geodesic in AdS$_3$ geometries may be phrased in terms of modular Berry phases. The volume of such a subregion is given in terms of the Crofton form on kinematic space \cite{Abt:2017pmf,Abt:2018ywl},
\begin{equation}
\frac{\operatorname{vol}(\Sigma)}{4 G_{N}^{2}}=\frac{1}{2 \pi} \int_{\mathcal{K}} \omega\left(\int_{\Delta_{p p^{\prime}}} \omega\right),
\label{eq:volume_subregion}
\end{equation}
where $\Delta_{p p^{\prime}}$ denotes the set of geodesics separating two bulk points $p$ and $p'$ and $\mathcal{K}$ refers to kinematic space. $\omega$ in \eqref{eq:volume_subregion} is identical to the Crofton form defined in this paper only for global AdS. For thermal AdS geometries, $\omega$ as specified in \cite{Abt:2017pmf,Abt:2018ywl} includes non-minimal geodesics, i.e. their kinematic space is composed of all geodesics rather than the shortest ones. Therefore, in the subregion complexity definition \eqref{eq:volume_subregion}, there are additional contributions unrelated to the entanglement entropy in thermal AdS geometries. For the vacuum, on the other hand, subregion complexity is defined entirely in terms of the Crofton form given only by the shortest geodesics and can hence be written in terms of the modular Hamiltonian. This seems to suggest that the CFT cost measure matching the result \eqref{eq:volume_subregion} for the vacuum may be defined in terms of the CFT modular Hamiltonian. It would be interesting to find the appropriate CFT cost measure for a CFT in the vacuum and see if the complexity measure can be adapted for thermal CFTs to account for additional contributions due to non-minimal geodesics in the dual bulk geometries. 

To conclude, we note that based on the results of this paper, there are many more interesting questions to be pursued in the future. For instance, it will be very interesting to connect our present construction to the split property in quantum mechanical systems and the lack thereof in gravitational systems as formulated extensively in \cite{Chowdhury:2021nxw, Raju:2021lwh}. It will also be extremely illuminating to understand the lack of factorization due to the non-exact symplectic form in terms of the emergent operator algebra of the dual CFT as discussed in a series of recent work \cite{Leutheusser:2021qhd, Leutheusser:2021frk, Witten:2021unn}. Finally, a further direction will be to consider multi-throat wormholes \cite{Emparan:2020ldj}.

\acknowledgments

We thank Ofer Aharony, Anton Alexeev,  Roberto Emparan, Marius Gerbershagen, Giuseppe Di Giulio, Kristan Jensen, Samir Mathur, Flavio Nogueira, Suvrat Raju, Julian Sonner, Jeroen van den Brink  and Suting Zhao for useful discussion.

We acknowledge support by the Deutsche Forschungsgemeinschaft (DFG, German Research Foundation) under Germany's Excellence Strategy through the Würzburg-Dresden Cluster of Excellence on Complexity and Topology in Quantum Matter - ct.qmat (EXC 2147, project-id 390858490). The work of A.-L.~W. is supported by DFG, grant ER 301/8-1 | ME 5047/2-1. The work of J.E. and R.M. is furthermore supported via the SFB 1170 `ToCoTronics' - project-id 258499086.

\appendix

\section{Modular Berry curvature for the vacuum}
\label{app:vacuum}
We briefly review how to obtain the modular Berry curvature for a CFT in the vacuum as calculated in \cite{Czech:2019vih}.

The modular Hamiltonian for a single interval $[u,v]$ employed in  \cite{Czech:2019vih}  is given by
\begin{equation}
H_{\mathrm{mod}}=K_++K_-=
s_{1} L_{1}+s_{0} L_{0}+s_{-1} L_{-1}+
t_{1} \bar{L}_{1}+t_{0} \bar{L}_{0}+t_{-1} \bar{L}_{-1},
\label{eq:modular_Hamiltonian_Czech}
\end{equation}
where \begin{equation}
\begin{aligned}
s_{1} &=\frac{2 \pi \cot \left(v-u\right) / 2}{e^{i v}+e^{i u}} & t_{1} &=-\frac{2 \pi \cot \left(v-u\right) / 2}{e^{i v}+e^{i u}} \\
s_{0} &=-2 \pi \cot \left(v-u\right) / 2 & t_{0} &=2 \pi \cot \left(v-u\right) / 2 \\
s_{-1} &=\frac{2 \pi \cot \left(v-u\right) / 2}{e^{-i v}+e^{-i u}} & t_{-1} &=-\frac{2 \pi \cot \left(v-u\right) / 2}{e^{-i u}+e^{-i v}}.
\end{aligned}
\end{equation}
In a vacuum CFT, every modular Hamiltonian may be mapped to another by a conformal transformation. Therefore, $P^{\lambda}_{0}(	\partial_\lambda H_{\mathrm{mod}})=0$ and \eqref{eq: equations modular Berry transport} reduces to $\partial_\lambda H_{\mathrm{mod}}=\left[V_{\delta \lambda}, H_{\mathrm{mod}}\right]$, which is solved by
\begin{equation}
V_{\delta u}=\frac{1}{2\pi i} \partial_{u}K_+.
\end{equation} 
Similarly, 
\begin{equation}
V_{\delta v}=-\frac{1}{2\pi i} \partial_{v}K_+.
\end{equation}
The Berry curvature operator for deformations $(u+\delta u, v+\delta v)$ is the obtained from \eqref{eq:Berry_curvature_Simple} and reads
\begin{equation}
\hat{R}_{ij}=-\frac{1}{2\pi i }\frac{K_+}{\sin^2\frac{v-u}{2}}.
\label{eq:modular_Berry_curvature_vacuum}
\end{equation}
It was observed that in \cite{Czech:2017zfq,Czech:2019vih} the Berry curvature for a CFT in the vacuum is related to the Crofton form, which reads
\begin{equation}
	\omega =\frac{1}{\sin^2\frac{v-u}{2}}du \wedge dv
\end{equation}
for the vacuum.
The analogy, however, can be taken one step further. As was shown in \cite{Penna:2018xqq}, the Crofton form on kinematic space can be identified with the symplectic form $\omega_{\mathrm{symp}}$ on the coadjoint orbit of $\frac{SO(2,1)}{SO(1,1)}\times\frac{SO(2,1)}{SO(1,1)}$. Therefore, 
\begin{equation}
\omega_{\mathrm{symp}}=	\omega_{\mathrm{Crofton}}.
\label{eq:relation_Crofton_symplectic_Berry_curvature}
\end{equation} 
This identifies the modular Berry curvature for the vacuum with the symplectic form on the vacuum coadjoint orbit of the global conformal group. Therefore, the modular Berry phase is a Berry phase on the coadjoint orbit of the global conformal group.

\bibliographystyle{JHEP}
\bibliography{bibliography_Berry.bib}
\end{document}